%% file: ms.tex
\documentclass[11pt]{article}
\pdfoutput=1
\usepackage{fullpage,url}
\usepackage{amssymb,amsbsy,amsfonts,amsmath,amsthm,xspace, mathrsfs}
\usepackage{graphicx,psfrag,caption,float,color,verbatim,subcaption, multirow}
\usepackage{changepage}
\usepackage{natbib}
\usepackage{authblk}
\usepackage[colorlinks,citecolor=blue,urlcolor=blue]{hyperref}

\newcommand{\F}[1]{\ensuremath{\mathrm{#1}}\xspace}

\begin{document}

\title{Graph-aware Modeling of Brain Connectivity Networks}
\author[a]{Yura Kim}
\author[a,b]{Daniel Kessler}
\author[a]{Elizaveta Levina}
\affil[a]{Department of Statistics, University of Michigan}
\affil[b]{Department of Psychiatry, University of Michigan}

\maketitle

\begin{abstract}
Functional connections in the brain are frequently represented by weighted networks, with nodes representing locations in the brain, and edges representing the strength of connectivity between these locations.  One challenge in analyzing such data is that inference at the individual edge level is not particularly biologically meaningful; interpretation is more useful at the level of so-called functional systems, or groups of nodes and connections between them; this is often called ``graph-aware'' inference in the neuroimaging literature.    However, pooling over functional regions leads to significant loss of information and lower accuracy.  Another challenge is correlation among edge weights within a subject, which makes inference based on independence assumptions unreliable.  We address both these challenges with a linear mixed effects model, which accounts for functional systems and for edge dependence, while still modeling individual edge weights to avoid loss of information. The model allows for comparing two populations, such as patients and healthy controls, both at the functional regions level and at individual edge level,  leading to biologically meaningful interpretations. We fit this model to  resting state fMRI data on schizophrenic patients and healthy controls, obtaining interpretable results consistent with the schizophrenia literature. 
\end{abstract}

\input{sections/intro}
\input{sections/methodsAOAS}
\input{sections/resultsAOAS}

\input{sections/discussion}

\section*{Acknowledgments}
\label{sec:acknowledgments}   
This research was supported in part by NSF DMS grants 1521551, 1646108, and 1916222, ONR grant N000141612910, a Rackham Predoctoral Fellowship from the University of Michigan awarded to D.\ Kessler, and a Dana Foundation grant to E. Levina, as well as by computational resources and services provided by Advanced Research Computing at the University of Michigan, Ann Arbor. We thank Prof.\ Stephan Taylor (Psychiatry, University of Michigan) and Prof.\ Chandra Sripada (Psychiatry and Philosophy, University of Michigan) and members of both of their labs for many useful discussions, and the Taylor lab for providing processed schizophrenia data. We thank Jesús Arroyo Relión (Statistics, Texas A\&M University) for his help with the data.

\bibliographystyle{agsm}
\bibliography{references}

\end{document}

%% file: sections/intro.tex
\section{Introduction}
\label{sec:intro}

Networks have been frequently used as a model for the brain's structural or functional connectome.    The types of nodes and edges depend on the data collection modality; we focus on data collected from functional magnetic resonance imaging (fMRI), although the statistical models we propose are applicable to other forms of brain imaging and potentially to other network data settings, particularly those involving multiplex networks, i.e., multiple networks observed on a common node set.     In brief, fMRI is obtained by recording blood oxygenation level dependent (BOLD) signals from subjects over time, at multiple locations in the brain;  the raw data for each subject is thus a 4-dimensional array (BOLD signal indexed by three spatial coordinates and time).    When extracting a network from fMRI data, a node is typically taken to be either a single location in the brain (a voxel) or a spatially contiguous group of voxels, otherwise known as a region of interest, or ROI \citep{Zalesky2010, Smith2012}.      Edges in brain networks capture connections between nodes, which can reflect either structural or functional connections depending on the type of data collected \citep{Bullmore2009, Bullmore2011}.  Structural connectivity has anatomical origins and can be inferred from fiber tracking methods such as diffusion MRI \citep{Zalesky2010, Zalesky2012, Craddock2013}.     Since we are working with fMRI data,  we focus on functional  connectivity, which represents temporal correlations between different parts of the brain \citep{Friston1994, VandenHeuvel2010}.     However, these two types of connectivity are frequently linked \citep{VanDenHeuvel2009}, and our methods are equally applicable to both types.    To take advantage of all the information available, we work with signed, weighted, dense networks where each pair of nodes is associated with a distinct real number; in contrast to some previous methods, e.g., \citet{Simpson2015}, we do not apply thresholding to convert this matrix of weights into a binary network.

Multiple studies \citep{Power2011, Yeo2011}  have produced brain atlases suggesting a robust parcellation of brain ROIs into functional systems, though details vary \citep{VandenHeuvel2010, Craddock2013}.     Many neurological and psychiatric disorders have been associated with changes in functional connectivity between such systems \citep{Craddock2009, Bullmore2011, Bullmore2012, Craddock2013}.   For instance,  in schizophrenia, a decrease in connectivity  between the frontal and temporal cortices has been reported \citep{Friston1995}.    The statistical challenge here is that while hypotheses and interpretation are framed at the level of connections between and within functional systems, the data are collected and modeled at a finer resolution of edges between ROIs. 
The scientific questions are often amenable to two-sample inference, comparing patients to healthy controls while identifying and locating specific changes in functional connectivity associated with a given disorder.

While the regression-based framework we propose is appreciably more general and can accommodate continuous covariates, the data we work with in this paper fall into the two-sample setting, containing  resting state fMRI data from 54 schizophrenic patients and 70 healthy controls. A more thorough description of the study and imaging parameters is available in \citet{Aine2017}.     ``Resting state'' fMRI involves imaging participants who are told to just relax in the scanner without being given any specific task to perform.    As a result, the per-voxel fMRI time series cannot be meaningfully temporally aligned between participants because there is no synchronized task or stimulus, and it is only meaningful to consider information averaged over time (after appropriate pre-processing; see \citet{Jesus2017} for details).    Resting state data are especially well suited for network-based analysis, since raw time series data cannot be meaningfully compared between patients, but connectivity networks can.

Much of the work on the problem of two-sample inference for brain connectivity networks falls into one of several categories as reviewed in \citet{chungStatisticalConnectomics2021} which differ chiefly in which features of the networks are to be compared.
The ``bag of (graph-theoretic) features'' framework uses a few global network summary measures such as modularity or the clustering coefficient and compares them across samples \citep{Bullmore2009}.
In this vein, some recent work by \citet{FUJITA201776} uses the spectral radius of each (thresholded) network to compare samples, considering, for example, whether the spectral radii of two sub-networks of a larger network are empirically correlated in a sample of networks.
Another common approach, sometimes referred to as ``bag of edges'' or ``massively univariate,'' arranges all the network edge weights into one vector, ignoring the network structure but allowing for usual multivariate inference at the edge level.   This approach requires correcting for massive multiple testing, which tends to be overly conservative when test statistics are correlated, as is the case here \citep{Craddock2013}.
In addition, some approaches focus on a more global comparison in the so-called ``bag of networks'' framework; for example \citet{tangSemiparametricTwoSampleHypothesis2017} proposes a test to evaluate whether two networks were drawn from the same generative model.
In contrast to these approaches, our method proposed below offers multi-resolution inference that can operate on both local edge-level features as well as more intermediate system-level features.


To improve power, {\citet{Zalesky2010} proposed a network-based statistic (NBS) approach, which reduces the number of multiple comparisons by focusing on large connected sub-networks.  
  Specifically, they consider the size of  the largest connected component of the graph obtained by retaining the edges with two-sample test statistic exceeding a given threshold, and compare to a null permutation-based distribution. However, the result depends on the threshold for the test statistic \citep{Kim2014}.
 \citet{belilovskyTestingDifferencesGaussian2016} also aim to exploit structure by assuming sparsity in the group differences: this method has the advantage of using the time courses directly but does not account for multiple subjects in each group.   Another set of methods uses  logistic regression with the binary group indicator as response and the individual's edge weights as explanatory variables to obtain various likelihood-based scores further used as a test statistics, with permutation-based $p$-values.   These include the sum of powered scores test \citep{Pan2014} and weighted and adaptive variants; for a detailed review and comparisons, see \citet{Kim2014}.  

A more recent algorithm proposed by \citet{Narayan2015} directly uses fMRI time series to test the hypothesis that the probability of each edge is the same for the two populations. \citet{Narayan2016} extended this method, called $R^3$ (resampling, random penalization, and random effects), to testing any (discrete or continuous) covariate effects, using  
random effects to account for  between- and within-subject variability.   
Mixed effects models generally have been gaining popularity for brain networks, because they allow for individual heterogeneity 
and provide a framework for testing covariate effects. 
For example, \citet{Sobel2014} use a linear mixed model for causal inference using fMRI time series data, though this work is not designed for two-sample inference.   

None of the methods discussed so far incorporate brain systems structure in the ROIs, but there is some recent work that leverages network community structure in modeling the association between brain connectivity and various phenotypes.
\Citet{xiaMultiscaleNetworkRegression2020} models each participant's brain connectivity as the sum of two low-rank matrices, one capturing the population mean and the other reflecting the contribution of covariates and their interactions at the level of functional systems.     We will compare this method to ours in Section \ref{sec:data}.

Our approach to modeling brain networks inherits all the advantages of a linear mixed effects model (binary or discrete covariates, individual effects, flexible variance structure) while accounting for network structure and enabling inference at both the system and the edge levels.   This allows for more accurate inference than what one can obtain by treating edges as a bag of features.    We incorporate system structure through a brain parcellation into functional systems.  ROIs within the same system of a meaningful parcellation tend to have  similar connectivity patterns  \citep{Smith2013}, and we leverage this property to parameterize the model in a more interpretable and concise fashion.   We also allow for some edge dependence induced by the parcellation, which leads to a more accurate assessment of uncertainty. 

Of course, implicit in our approach are various assumptions, e.g., on the form of the variance structure as discussed in Section \ref{sec:graph-aware-variance}.
While these assumptions are unlikely to be perfectly satisfied, in line with the classic wisdom of \citet{box-allmodels}, we believe that our model may be wrong yet useful.
For example, as we see in Section \ref{sec:data}, our approach offers far more accurate uncertainty quantification, and thus more valid inference, than a more naive approach.
In addition, conducting inference at the level of the brain system better aligns with prevailing scientific thinking and enables the confirmation of existing scientific insights as well as the identification of potentially novel effects.

A particularly related line of work on  mixed effects models for brain networks was initiated by \citet{Simpson2015} and continued in \citet{Bahrami2017,Bahrami2019,Simpson2019};  we will collectively refer to this body of work as the S-L (Simpson-Laurienti) approach.      There are several important differences between the S-L approach and ours.  First,  S-L estimates a relatively small number of global coefficients whereas we have both system-level and edge-level parameters, allowing for greater model flexibility. 
Second, we directly model edge weights without thresholding, while S-L fits a two-stage model, with first stage determining a subset of important edges and the second stage only modeling those. 
Finally, the covariance structure of our model is appreciably more general, including non-zero covariance between both random effects and residuals.  
Of course, if the simpler assumptions of the S-L approach hold, their model may have both computational and power advantages, but such assumptions are unlikely to be verifiable in practice.  

Another related paper by \citet{FIECAS2017256} proposes a mixed effects model somewhat similar to ours for two-sample testing at either the level of the entire network or for individual edges.   In contrast to our approach, they conduct two-sample inference by fitting two separate models and then comparing resulting statistics instead of testing parameters corresponding to group differences within a single model.
They model within-subject covariance using a non-parametric approach based on sampling distributions of correlations under autocorrelation, combined with somewhat restrictive assumptions about remaining error structure.    This works well for their setting of 11 ROIs but they note the approach does not scale well, whereas our covariance model is easily applicable to our dataset with 
$235$ ROIs.

Next, we present  the graph-aware linear mixed model for brain networks and the fitting algorithm, in Section \ref{sec:methods}. Section \ref{sec:data} presents empirical results including analysis of the COBRE dataset, and Section \ref{sec:discussion} concludes with discussion and possible directions for future work.



%% file: sections/methodsAOAS.tex
\section{Statistical methods:  a network-aware mixed effects model}
\label{sec:methods}

\subsection{Setup and notation}

We assume that we are given a sample of $N$ networks on $n$ nodes; each network is represented by its weighted $n \times n$  adjacency matrix $A_{m}$, $m=1, \ldots ,N$.    The nodes are aligned across all networks, corresponding to a common ROI atlas in the brain application.   The entry $A_{m, ij}$ represents the connectivity between nodes $i$ and $j$ for network $m$, and we focus on the undirected setting $A_{m, ij}=A_{m, ji}$ with no self-loops, appropriate for fMRI data.   
In the brain application, the weights are Fisher-transformed Pearson correlations between time series at different ROIs, which are standard in the neuroimaging literature.  Alternative measure of connectivity, such as partial correlations or thresholded correlations, can also be used; see  \citet{Zhen2007, Smith2011, Liang2012, Craddock2013} for discussions on various ways of measuring functional connectivity.     There are strong local spatial correlations between the time series at neighboring ROIs, which leads to highly dependent edge weights.

We assume that the network nodes are divided into groups corresponding to network communities; in the application, this corresponds to ROIs grouped into functional systems.    In network analysis, communities are typically viewed as groups of nodes with similar connectivity patterns;  in many cases, including typical brain networks, this means that there are stronger connections within communities than between them.   In this paper, we use an existing known parcellation of the brain into functional systems;  alternatively, one could apply one of the many community detection techniques first to estimate such a parcellation.

Let $c_i$ be the community label of node $i$, common across all networks and taking values in $\{1, \dots, K\}$.   We refer to an unordered pair of communities $(a,b)$, where $a, b \in \{1, \dots, K\}$ as a network cell;  there are a total of $K(K+1)/2$ cells
corresponding to $K$ communities.
We will use these cells as a target of inference when characterizing effects at the cell-level and the cells will also inform our covariance structure as discussed in Section \ref{sec:graph-aware-variance}.

Let $n^{(a)}=\lvert  \left\{ i: c_i= a \right\} \rvert$ denote the number of nodes in community $a$, and let $n^{(a, b)}$ be the number of edges in cell $(a, b)$, where
\begin{equation*}
n^{(a, b)}=\begin{cases}n^{(a)}n^{(b)},&\mbox{if } a\neq b\\
n^{(a)}(n^{(a)}-1)/2& \mbox{if } a=b,
\end{cases}
\end{equation*}
and let $n^{(\cdot, \cdot)} = n(n-1)/2$ be the total number of distinct edge weights.

Next, let $y^{(a, b)}_m \in \mathbb{R}^{n^{(a, b)}}$ be the vector of edge weights in cell $(a, b)$ for subject $m$.    We use $y^{(a, b)}_{m, i}$ to refer to an element of these vectorized edge weights, with $i$ ranging from $1$ to  ${n^{(a, b)}}$ for cell $(a,b)$.  Finally, let $y_m \in \mathbb{R}^{n^{(\cdot, \cdot)}}$ be the vector of \emph{all} edges for subject $m$, i.e., the concatenation of $y^{(a, b)}_m$ across all cells, and $y \in \mathbb{R}^{N n^{(\cdot, \cdot)}}$ the vector of all edges for all subjects.

\subsection{A Linear Mixed Effects Model}

Brain connectivity naturally varies across subjects, even if they have the same disease status and other covariates.   A mixed effects model is a natural tool to incorporate this individual variation while also modeling effects of other covariates.  A linear model allows for a straightforward interpretation of these effects, and can account for the high correlations among edge weights by including a general covariance error structure.    

  For simplicity, we first write out the model for a given network cell $(a,b)$.
Let $x_m \in \mathbb{R}^p$ be a vector of subject-level covariates (generally including a ``1'' term for the intercept), such as disease status, age, gender, and so on.
For each cell $(a, b)$, we partition covariate effects into cell-level effects, denoted by coefficients $\alpha^{(a, b)} \in \mathbb{R}^p$, and additional edge-level effects, denoted by coefficients  $\eta_i^{(a, b)} \in \mathbb{R}^p$, for each edge $i$ in cell $(a,b)$.  We then model the expected edge weight as 
\begin{equation*}
\mathbb{E} \left(y^{(a, b)}_{m, i} \right) = x_m^{T} \left( \alpha^{(a, b)}+ \eta^{(a, b)}_i \right),
\end{equation*}
where $m=1, \dots,N$ is the subject index, $i = 1, \dots, n^{(a, b)}$ is the edge index within the cell, and $a, b \in \{1, \dots K\}$ are community labels.
For identifiability, we require that $\displaystyle \sum_i \eta^{(a, b)}_{i, j}=0$ for all $j$.

Adding a  cell-specific subject random effect term $ \gamma_{m}^{(a, b)}$ and an error term $\epsilon_{m, i}^{(a, b)}$, we get the proposed linear mixed effects model for edge weights,  
\begin{equation*}
y_{m, i}^{(a, b)}=   x_m^{T}\alpha^{(a, b)}+x_m^{T}\eta^{(a, b)}_i + \gamma_{m}^{(a, b)}+\epsilon_{m, i}^{(a, b)} \ . 
\end{equation*}
The noise variables $\epsilon_{m, i}^{(a, b)}$, specific to each subject and each edge, have mean 0 and are independent of the random effects. 

As an example, consider the two-sample setting  where we  have a single subject covariate, say $d_m$, which is an indicator of, for example, disease status of subject $m$, set to $1$ if subject $m$ has the disease and $0$ otherwise.   The intercept then represents the global cell mean of subjects who do not have the disease.   In this case, we have $x_m = \begin{pmatrix} 1 & d_m \end{pmatrix}^T$, and the model becomes 
\begin{equation*}
y_{m, i}^{(a, b)}=\left( \alpha^{(a, b)}_0+d_m\alpha^{(a, b)}_1 \right) + \left(\eta^{(a, b)}_{i, 0}+d_m\eta^{(a, b)}_{i, 1} \right)+\gamma_{m}^{(a, b)}+\epsilon_{m, i}^{(a, b)} \ . \\
\end{equation*}
For every cell $(a,b)$, the term $\alpha_0$ represents the cell-level mean for patients with no disease,  $\alpha_1$ the cell-level shift due to disease, $\eta_{i, 0}$ the edge-specific intercept for patients with no disease, and $\eta_{i, 1}$ the edge-specific disease effect.
These are all fixed effects, whereas $\gamma_m^{(a, b)}$ is the subject-specific random effect for the given network cell representing individual heterogeneity, with mean 0 over the population of subjects.

\subsection{Modeling edge dependence}

While we have now set up a mean model for each  network cell $(a,b)$,  there are correlations among edge weights across the whole brain.    Not modeling these correlations is inaccurate and will result in overly optimistic estimates of uncertainty \citep{Li2015};   modeling all of them will result in an unmanageable number of parameters, so a compromise is needed.   

First, we rewrite the model collecting the terms for all edges together.
Recall that $y_m$ is the vector of all edges for subject $m$, and let $\gamma_m \in \mathbb{R}^{K ( K + 1)/2}$ be the vector collecting all cell-level random effects for subject $m$.   We assume that each random effects vector $\gamma_m$ has mean 0 and covariance matrix $U$, and the edge weights satisfy
\begin{equation}
\mathbb{E} \left(y_{m} \mid \gamma_{m} \right)=X_{m}\beta+Z\gamma_{m},\ \ \mathbb{V}\mbox{ar} \left(y_{m}\mid \gamma_{m} \right)=V
\label{eq:model_conditional}
\end{equation}
where $\beta$ is a vector that captures the contribution of the coefficients in $\alpha$ and the $\eta_i$'s for all cells (one can easily move between unconstrained $\beta$ and the original constrained parameterization).
The random effect design matrix is responsible for ``broadcasting'' the cell-level effects $\gamma_m^{(a, b)}$ across edges and has the block-diagonal form
\begin{align*}
  Z =
  \begin{pmatrix}
    1_{n^{(1, 1)}} & 0_{n^{(1, 1)}} & 0_{n^{(1, 1)}} & \ldots & 0_{n^{(1, 1)}}\\
    0_{n^{(2, 1)}} & 1_{n^{(2, 1)}} & 0_{n^{(2, 1)}} & \ldots & 0_{n^{(2, 1)}}\\
    0_{n^{(2, 2)}} & 0_{n^{(2, 2)}} & 1_{n^{(2, 2)}} & \ldots & 0_{n^{(2, 2)}} \\
    \vdots & \vdots & \vdots & \ddots & \vdots \\
    0_{n^{(K, K)}} & 0_{n^{(K, K)}} & 0_{n^{(K, K)}} & \ldots & 1_{n^{(K, K)}}
  \end{pmatrix},
\end{align*}
where $1_{n^{(a, b)}}$ and $0_{n^{(a, b)}}$ are a vector of either all ones or all zeroes with length $n^{(a, b)}$.
The fixed effects design matrix for cell $(a, b)$ is given by 
\begin{equation*}
  X^{(a, b)}_{m} =
  \begin{pmatrix}
    x_m^T & x_m^T &  \ldots  & 0 \\
    x_m^T & 0 &  \ldots  & 0 \\
    \vdots & \vdots & \ddots & \vdots \\
    x_m^T & 0 &  \ldots  & x_m^T \\
    x_m^T & -x_m^T &  \ldots  & -x_m^T
  \end{pmatrix},
\end{equation*}
where the initial columns hold the cell-level effects $\alpha^{(a, b)}$, and subsequent columns capture the contribution of the $\eta_i^{(a, b)}$'s.
The full design matrix for subject $m$ is then given by ``tiling'' cell-level design matrices into a block-diagonal $X_m = \operatorname{diag} \left( X_m^{(a, b)} \right)$. 
Integrating out the random effect gives
\begin{equation}
\mathbb{E} \left( y_{m} \right)=X_{m}\beta,\ \ \mathbb{V}\mbox{ar}\left( y_{m} \right)=V+ZUZ^{T} \equiv \Sigma.
\label{eq:model_marginal}
\end{equation}

We assume that subjects are not correlated, in which case the overall covariance of $y$ is given by a block diagonal matrix with repeated blocks of $\Sigma$ on the diagonal and 0 elsewhere, and the overall design matrix $X = \begin{pmatrix} X_1^T & X_2^T & \ldots & X_N^T \end{pmatrix}^T$.
This block structure in the covariance allows us to avoid direct inversion of a very large matrix:  assuming all inverses are well defined, the best linear unbiased estimator of $\beta$ is given by the generalized least squares (GLS) estimator, 
\begin{equation}
\hat{\beta}=\left( \sum_{m} X_{m}^{T} \Sigma^{-1} X_{m} \right)^{-1} \left(\sum_{m} X_{m}^{T} \Sigma^{-1} y_{m} \right).
\label{eq:gls}
\end{equation} 
In fact all estimators of this form, even when the assumed covariance structure $\check{\Sigma} \neq \mathbb{V}\mbox{ar}\left( y_{m} \right)$, 
are unbiased estimators of $\beta$ under model \eqref{eq:model_marginal}, since
\begin{align*}
  \mathbb{E} \left( \check{\beta} \right)
  &= \left( \sum_{m} X_{m}^{T} \check{\Sigma}^{-1} X_{m} \right)^{-1} \left( \sum_{m} X_{m}^{T}\check{\Sigma}^{-1}\mathbb{E} \left( y_{m} \right) \right) 
=\beta .
\end{align*}

\subsection{Graph-aware variance structure}
\label{sec:graph-aware-variance}
There are multiple reasons to impose structure on the variance, in addition to computational savings.    With no additional assumptions on $V$, the decomposition $\Sigma=V+ZUZ^{T}$ is not unique, and  $V$ and $U$ are not identifiable.    There are multiple options for solving this identifiability problem, and incorporating  network cell structure into variance assumptions, as we have done for the mean, has the additional benefit of being faithful to the data structure in the application.  

\begin{figure}[htb]
\centering
\includegraphics[width = 4in]{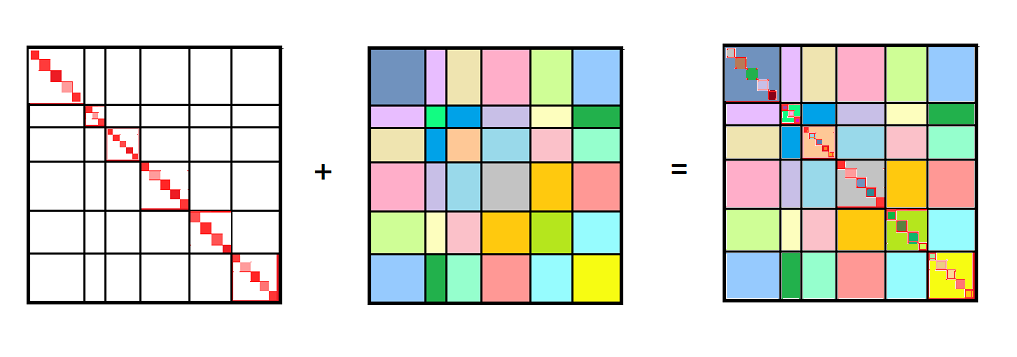}\\
\includegraphics[width = 4in]{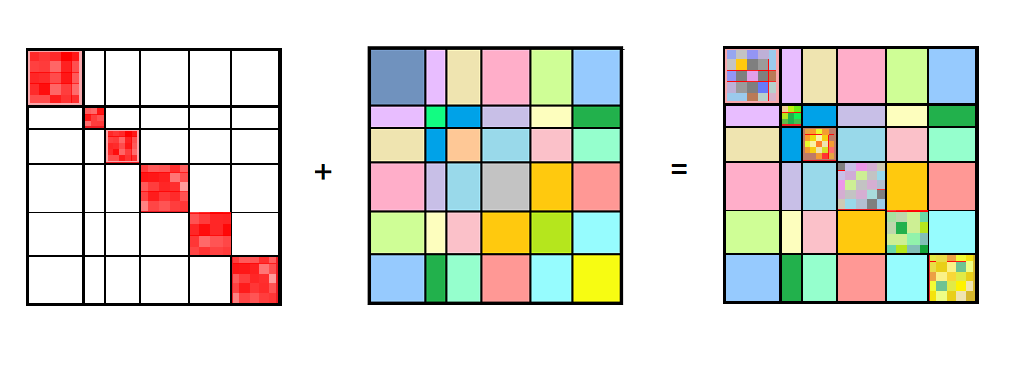}
\vspace{-0.5cm}
\caption{Decomposition of the intra-subject covariance matrix of edge weights, $V+ZUZ^{T} = \Sigma$ with diagonal $V$ (top)  and block-diagonal $V$ (bottom).}
\label{fig16}
\end{figure}

Recall that we impose network structure through using cells $(a,b)$.   In the decomposition $\Sigma=V+ZUZ^{T}$, the second term already has a block structure corresponding to network cells,  
and it would be natural to impose some structure on $V$ that creates a structure corresponding to network cells in $\Sigma$.  
Figure~\ref{fig16} shows two examples that achieve this goal:  a diagonal $V$ and a block-diagonal $V$.   
Diagonal $V$ allows for heteroscedastic noise at each edge (which we can estimate because we have multiple subjects), and block-diagonal $V$ further allows for dependence between edge noise variables belonging to the same network cell.    In both cases, the resulting covariance matrix $\Sigma = V + Z U Z^{T}$  is dense, allowing for dependence between all edge weights.  
 
\subsection{Model Fitting with the EM algorithm}

In practice, the GLS estimator \eqref{eq:gls} is not computable, since  $V$ and $U$ are unknown.  We take the approach of jointly estimating $V$, $U$, and $\beta$ by maximum likelihood under the normal assumption on the random effects and the errors.     Specifically, we assume that $\gamma_{m}  \sim  N(0,U)$ and 
$$
y_{m}  =  X_{m}\beta+Z\gamma_{m} + \epsilon_m, 
$$
where $\epsilon_m = \big\{\{\epsilon^{(a, b)}_{m, i}, 1\le i \le n^{(a, b)}\}: 1\le a \le b \le K\big\} \sim N(0,V)$ is the concatenated error term independent of $\gamma_m$.   

The normal assumption is reasonable for our edge weights measured by Fisher's $z$-transformation of the Pearson correlation coefficient $r$, which is designed to make the correlations close to normally distributed.   The transformation is defined as
\begin{equation*}
z=\frac{1}{2} \log \left(\frac{1+r}{1-r}\right),
\end{equation*}
and is commonly used in neuroimaging \citep{Varoquaux2013}.     Since normality is difficult to verify in practice, this can also be viewed as a generic $M$-estimation approach with a loss function corresponding to the normal likelihood.

We use the EM algorithm to find maximum likelihood estimates (MLE) of $V$, $U$, and $\beta$.   The derivation under the normal assumption is straightforward and is omitted here.  The algorithm consists of the following two steps, iterated until convergence once initialized.   For conciseness, we write $<f>$ to denote the conditional expectation of $f$ given the observed data $y$.

\subsubsection*{E-Step}   Calculate posterior means of $\gamma_{m}$ and $\gamma_{m}\gamma_{m}^{T}$ given the data, as 
\begin{equation*}
<\gamma_{m}> = UZ^{T}(V+ZUZ^{T})^{-1}(y_{m}-X_{m}\beta),
\end{equation*}
\begin{equation*}
<\gamma_{m}\gamma_{m}^{T}> = (U-UZ^{T}(V+ZUZ^{T})^{-1}ZU)+ <\gamma_{m}><\gamma_{m}>^{T}.
\end{equation*}

\subsubsection*{M-Step}  Update the estimate of $U$ to 
\begin{equation*}
\hat U = \frac{1}{N}\sum_{m}<\gamma_{m}\gamma_{m}^{T}>,
\end{equation*}
and update the estimates of $V$ and $\beta$ by repeating the following steps until convergence:
\begin{align*}
\hat V_0 & =  \frac{1}{N}\sum_{m}<(y_{m}-X_{m}\hat \beta-Z\gamma_{m})(y_{m}-X_{m}\hat \beta-Z\gamma_{m})^{T}> \\
\hat V & =\begin{cases}\mbox{diag}(\hat V_0),&\mbox{if } V \mbox{ is modeled as diagonal}\\
\mbox{block-diag}(\hat V_0), & \mbox{if } V \mbox{ is modeled as block-diagonal},
\end{cases} \\
  \hat \beta &  = \left(\sum_{m} X_{m}^{T} \hat V^{-1}X_{m}\right)^{-1} \sum_{m}X_{m}^{T}\hat V^{-1}(y_{m}-Z<\gamma_{m}>) .
\end{align*}

\subsubsection*{Initialization}  The coefficients  $\hat \beta$ can be initialized by ordinary least squares, taking 
\begin{equation*}
\hat \beta^{(a, b)} = \left(\sum_{m} \left( X^{(a, b)}_{m} \right)^{T} X^{(a, b)}_{m} \right)^{-1} \sum_{m} (X^{(a, b)}_{m})^{T}y^{(a, b)}_{m}, \ \ \hat \beta = \{\hat \beta^{(a, b)} : 1\le a\le b\le K\}.
\end{equation*}

To initialize $V$, we first calculate, for each pair of cells $(a, b), (c, d)$, the empirical covariance
\begin{equation*}
  \hat \Sigma^{(a, b), (c, d)}=\frac{1}{N-1}\sum_{m} \left( y_{m}^{(a, b)}-X_{m}^{(a, b)}{\hat \beta}^{(a, b)} \right) \left(y_{m}^{(c, d)}-X_{m}^{(c, d)}{\hat \beta}^{(c, d)} \right)^{T}.
\end{equation*}
Then, we initialize $U=(U^{(a, b), (c, d)})$ by taking
\begin{equation*}
\hat U^{(a, b), (c, d)}=  \frac{1}{n^{(a, b)}n^{(c, d)}}\sum_{i=1}^{n^{(a, b)}}\sum_{j=1}^{n^{(c, d)}}\hat\Sigma^{(a, b), (c, d)}_{ij} , 
\end{equation*}
and initialize $V$ with a diagonal matrix regardless of what assumptions we later make about it, as 
\begin{equation*}
\hat V^{(a, b)} = \left(\frac{1}{n^{(a, b)}}\sum_{i=1}^{n^{(a, b)}}\hat \Sigma^{(a, b), (c, d)}_{ii} - \hat U^{(a, b), (c, d)} \right)I_{n^{(a, b)}}, \ \ \hat V=\mbox{diag}(\hat V^{(a, b)}),
\end{equation*}
where $I_n$ is the $n\times n$ identity matrix.

\subsubsection*{Implementation}

To increase the stability of the algorithm and to speed up convergence, we implement the EM algorithm for the equivalent model:
\begin{equation*}
y_{m, i}^{(a, b)}=\zeta_{m}^{(a, b)}+x^T_m\eta_{i}^{(a, b)}+\epsilon_{m, i}^{(a, b)} .
\end{equation*}
Writing $\zeta_{m}=\{\zeta^{(a, b)}_{m}\}_{1\le a\le b\le K}$, $\zeta_{m}^{(a, b)} = x^T_m\alpha^{(a, b)}+\gamma_{m}^{(a, b)}$ and $\mu_m = \{x^T_m\alpha^{(a, b)}\}_{1\le a\le b\le K}$, we have 
\begin{equation*}
\zeta_{m} \sim N(\mu_m,U),
\end{equation*}
This simply combines the terms $x^T_m\alpha^{(a, b)}$ and $\gamma_{m}^{(a, b)}$, both depending only on $m$ and $(a, b)$, to get `mean-shifted' random effects terms, $\{\zeta_{m}^{(a, b)}\}$. This model is mathematically equivalent to the previous one, but empirically converges faster and is less dependent on the initial value of $\beta$, due to centering.  

The most time-consuming part of the algorithm is updating $\beta$ in the M-step, which involves inverting the large matrix
\begin{equation*}
\sum_{m}(X_{m}^{T}V^{-1}X_{m}).
\end{equation*}
In a typical neuroimaging application, the size of this matrix will be in the tens of thousands;  for the COBRE dataset analyzed in Section  \ref{sec:data}, it is approximately $28,000 \times 28,000$.  To avoid inverting this matrix,  we instead solve for $\beta$  using a block coordinate descent algorithm.   
This implementation takes around 5 minutes to fit the model to the COBRE data with a diagonal $V$ and around 15 minutes with a block-diagonal $V$ on a machine with ten 2.8 GHz Intel Xeon E5-2680v2 processors, each with 8GB of memory.
In contrast, the OLS estimator that we compare against is almost trivially fast to compute (less than 1 minute) and indeed is essentially solved during the initialization of our approach.



%% file: sections/resultsAOAS.tex
\section{Empirical Results}
\label{sec:data}


The COBRE schizophrenia dataset, introduced in Section \ref{sec:intro},  contains resting state fMRI connectivity brain networks of 54 patients with schizophrenia and 70 healthy controls.   
Data were downloaded via NITRC (\url{http://fcon_1000.projects.nitrc.org/indi/retro/cobre.html})
and processed by Prof. Stephan Taylor's lab in the Department of Psychiatry at the University of Michigan.     This dataset is also available via the COINS platform \citep{landis_coins_2016,wood_harnessing_2014}.

\begin{figure}[htb]
\centering
\includegraphics[width = 3in]{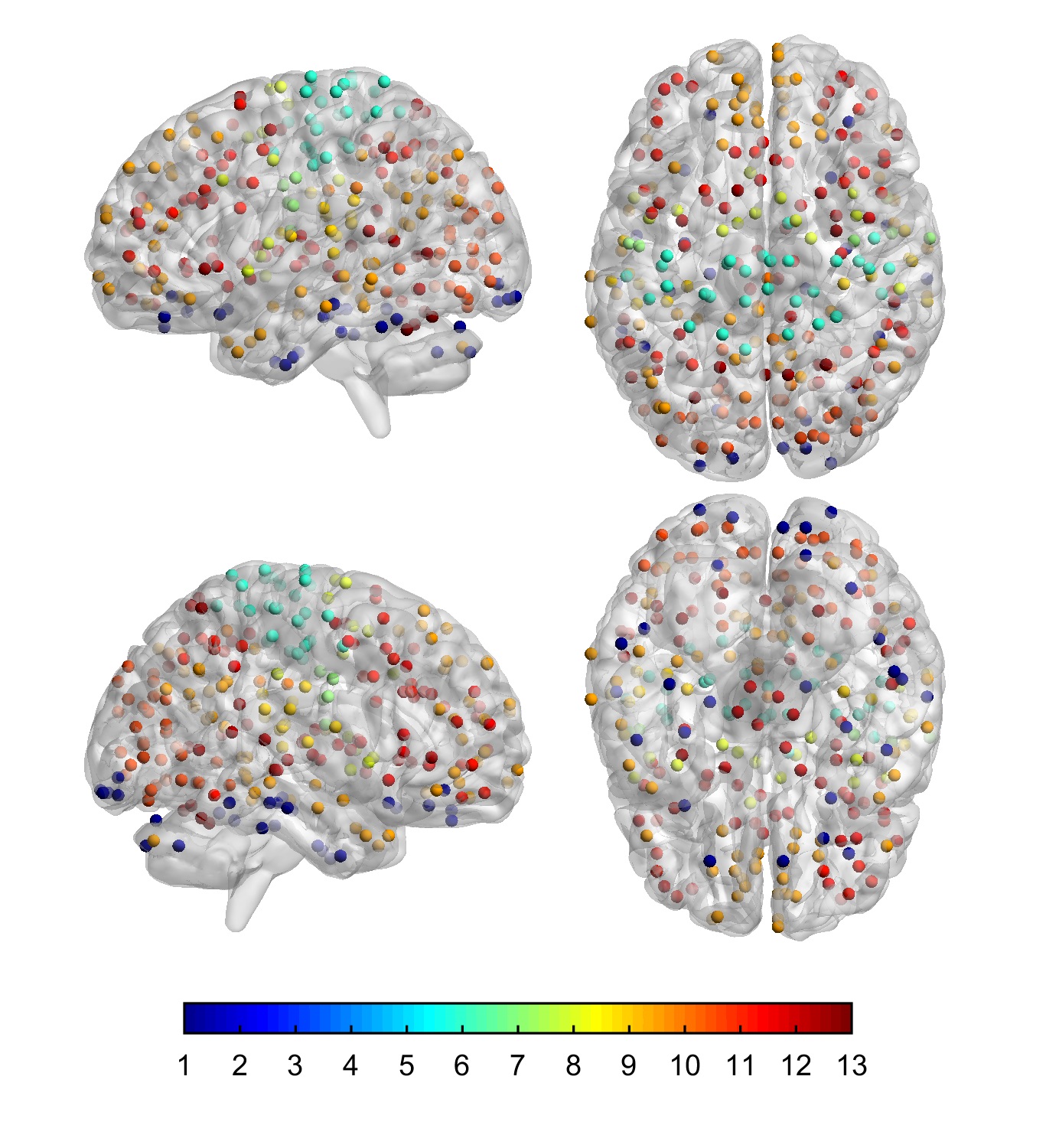}
\caption{The 13 functional systems  from \citet{Power2011} represented by color.
  Left column:  Sagittal view from the left (top) and the right
  (bottom). Right column:  axial view from above (top) and from below (bottom). Figure generated using BrainNet \citep{Xia2013}.}
\label{fig1}
\end{figure}

After pre-processing, individual voxels are combined through spatial smoothing into 264 ROIs from  the functional parcellation by \citet{Power2011}, and Fisher-transformed pairwise Pearson correlations between the time series at each of the ROIs are used as edge weights.     Empirically, these weights are approximately normal, which is expected from Fisher-transformed correlations.    The parcellation of \citet{Power2011} divides the 264 ROIs into 14 functional systems, which play the role of communities;  these systems are shown in Table~\ref{table1} and Figure~\ref{fig1}. We used only systems 1 through 13, and excluded the 28 nodes of the ``Uncertain'' system from the analysis, since we have no a priori reason to believe that nodes that could not be clearly assigned to any system have a homogeneous connectivity pattern.    Also, the data for node 75 are missing from the COBRE dataset, which leaves a total of $264-28-1=235$ ROIs for the subsequent analysis.
In addition, as part of our collaborator's pre-processing pipeline, nuisance covariates including age, gender, motion (summarized as mean framewise displacement and its square), and handedness were removed before fitting our model.
This processed data was also used in \citet{Jesus2017} and is available in the \href{https://github.com/jesusdaniel/graphclass}{graphclass package}.

We use this dataset as a basis for two experiments where we can control ground truth:
in Section \ref{sec:semi-synthetic-experiment}, we assess the performance of our methods on synthetic data drawn from a model based on a fit to the COBRE data, while in Section \ref{sec:hc-only-analysis} we assess the validity of our methods under the global null by using only randomly labeled healthy controls.
We next fit our method to the full COBRE dataset and present the estimated parameters in Section \ref{sec:cobre-fitting}, conduct inference in Section \ref{sec:cobre-inference}, and then apply a related, competing method in Section \ref{sec:cobre-compare-xia}.

\begin{table}[ht]
\footnotesize
\centering
\begin{tabular}{rlr}
  \hline
 & System & Number of nodes   \\ 
  \hline
1 & Sensory/somatomotor Hand & 30  \\ 
2 & Sensory/somatomotor Mouth & 5  \\ 
3 & Cingulo-opercular Task Control & 14 \\
4 & Auditory & 13 \\
5 & Default mode & 58 \\ 
6 & Memory retrieval & 5 \\ 
7 & Visual & 31 \\
8 & Fronto-parietal Task Control & 25 \\
9 & Salience & 18 \\ 
10 & Subcortical & 13 \\ 
11 & Ventral attention & 9 \\
12 & Dorsal attention & 11 \\
13 & Cerebellar & 4 \\ 
-1 & Uncertain & 28 \\ 
   \hline
\end{tabular}
\caption{Functional systems from \citet{Power2011}.} 
\label{table1}
\end{table}

\subsection{Assessing the effect of variance modeling in synthetic data}
\label{sec:semi-synthetic-experiment}

Before we proceed to analyze the COBRE data, we perform a comparison
of several versions of our method on synthetic data simulated based on
the COBRE dataset,  but in a way that allows us to vary parameters of interest.  The goal is to understand the effect of variance modeling on discovering individual effects of interest in a realistic setting where we nonetheless know the truth.   To generate synthetic data, we first fitted the linear mixed effects model \eqref{eq:model_conditional} to the COBRE dataset using a diagonal matrix $V$. Then, for each network cell $(a,b)$,  we calculated the $p$-value for the  $z$-test of the hypothesis $H_0:\alpha^{(a, b)}_1=0$, 
where $\alpha^{(a, b)}_1 $ is the coefficient of the binary disease indicator $x_{m}$ in cell $(a,b)$ ($x_{m}= 0$ for healthy controls and 1 for schizophrenic patients).    Standard errors were computed from the standard GLS variance estimator, taking the square root of the diagonal elements of
\begin{equation*}
\left( \sum_{m} X_{m}^{T} \left( V + ZUZ^{T} \right)^{-1}X_{m} \right)^{-1}.
\end{equation*}
We chose as ``true positives'' the 13 cells with $p < 0.05$, and set $\alpha_1^{(a, b)} = 0$ for the remaining 78 cells for the purpose of simulating synthetic data.
Using this information, we refitted the GLS model with EM to obtain a diagonal $\hat{V}$, $\hat{U}$, and $\hat{\beta}$ (which comprises both cell- and edge-specific effects), with the 78 true negative $\alpha_1$'s set to 0.  Then we generated a new synthetic dataset by drawing
\begin{equation*}
y_{m} \sim N(X_{m}\hat{\beta},\hat{V}+Z\hat{U}Z^T),\ m=1, \ldots ,100, 
\end{equation*} 
with 50 subjects each from healthy and schizophrenia populations. 
Our goal was to compare the usual OLS estimator 
\begin{equation*}
\hat{\beta}^{OLS}=\left(\sum_{m} X_{m}^{T}X_{m} \right)^{-1} \left(\sum_{m} X_{m}^{T}y_{m} \right),
\end{equation*}
to the proposed GLS estimator.    For each generated dataset, we fitted our model, with all 91 covariates, in two ways, with  either diagonal or block-diagonal $V$, and we also fitted OLS for comparison.   For each estimator, we computed standard errors for GLS and OLS according to their respective standard formulas.     The entire simulation was repeated 100 times.   

While we fitted GLS using both a diagonal $V$ and block-diagonal $V$ covariance structure, the model from which we drew data \emph{perfectly} coincides with the former.
While the block-diagonal $V$ covariance structure contains diagonal $V$ as a special case, it is appreciably more flexible, and we anticipated that this additional flexibility may result in some overfitting in this case, although as we shall see the effect is quite modest.

\begin{figure}[H]
\centering
\begin{subfigure}{1\textwidth}
  \centering
      \caption{OLS}  \vspace{-0.2cm}
  \includegraphics[width = 5in]{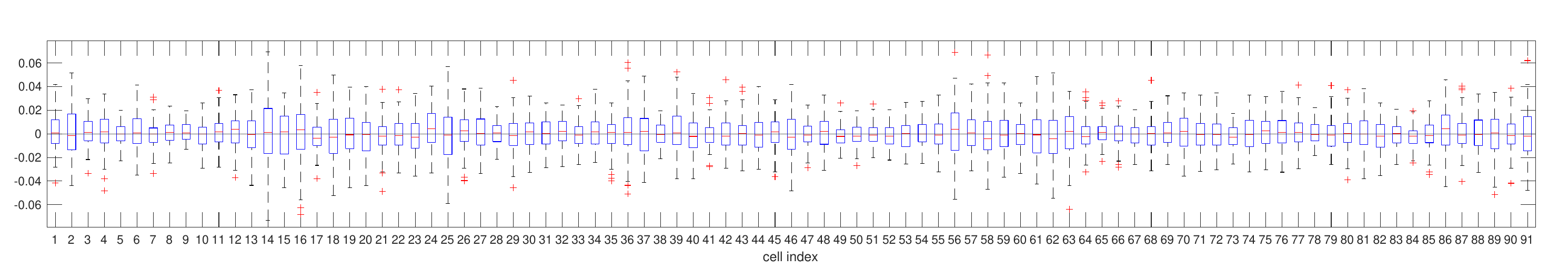}
\end{subfigure}%
\\
\begin{subfigure}{1\textwidth}
  \centering
   \caption{GLS, diagonal $V$}   \vspace{-0.2cm}
  \includegraphics[width = 5in]{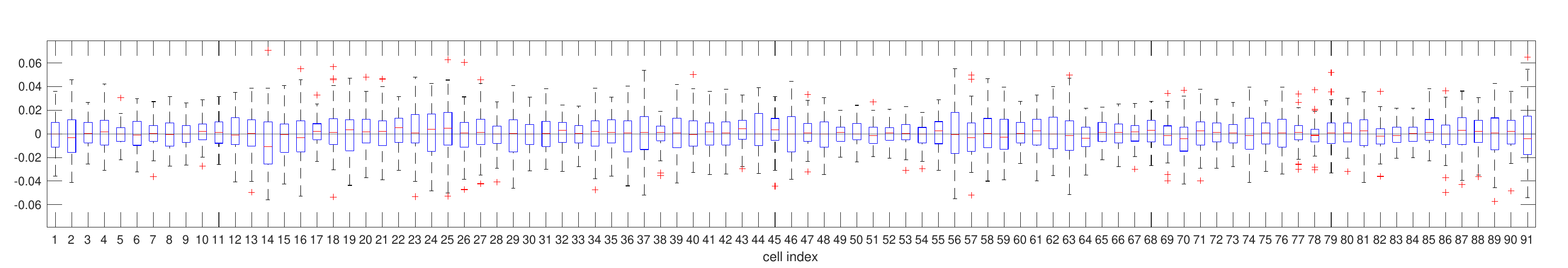}
\end{subfigure}%
\\
\begin{subfigure}{1\textwidth}
  \centering
   \caption{GLS, block-diagonal\ $V$}   \vspace{-0.2cm}
  \includegraphics[width = 5in]{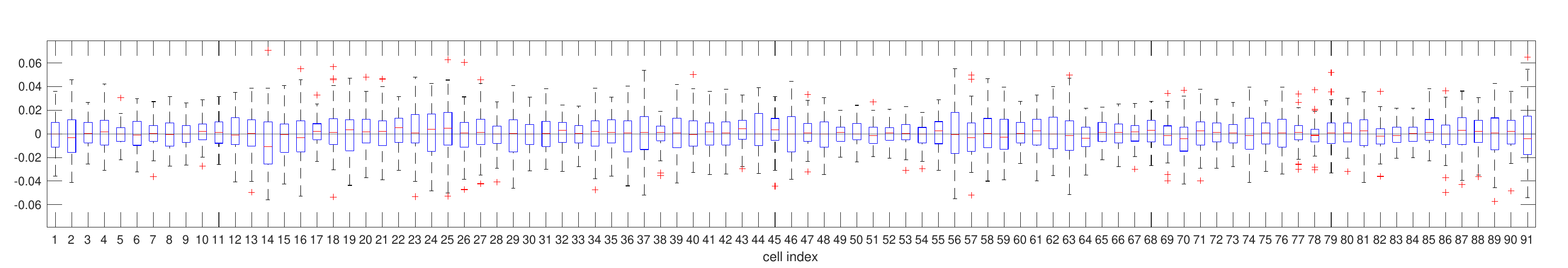}
\end{subfigure}%
\vspace{-0.1cm}
\caption{Boxplots of the errors, $\hat{\alpha}_1-\alpha_1$, with $100$ replications.
  Outliers are indicated by $+$'s, and a horizontal line marks $0$ (no error).
}
\label{fig6}
\end{figure}

Figure~\ref{fig6} shows the boxplots of errors for the main parameter of interest, cell-level difference between populations $\hat{\alpha}_1-\alpha_1$, for the three types of estimators and the 91 network cells.     All three estimators look similar, and all the boxplots are centered around 0, as they should be since all three estimators are unbiased.
A perhaps surprising result is that the point estimates for both diagonal and block-diagonal $V$ methods are precisely the same.
This is a consequence of looking at only effects at the level of network cells and of the particular covariance structures assumed.

\begin{figure}[H]
\centering
\begin{subfigure}{1\textwidth}
  \centering
      \caption{OLS}  \vspace{-0.2cm}
  \includegraphics[width = 5in]{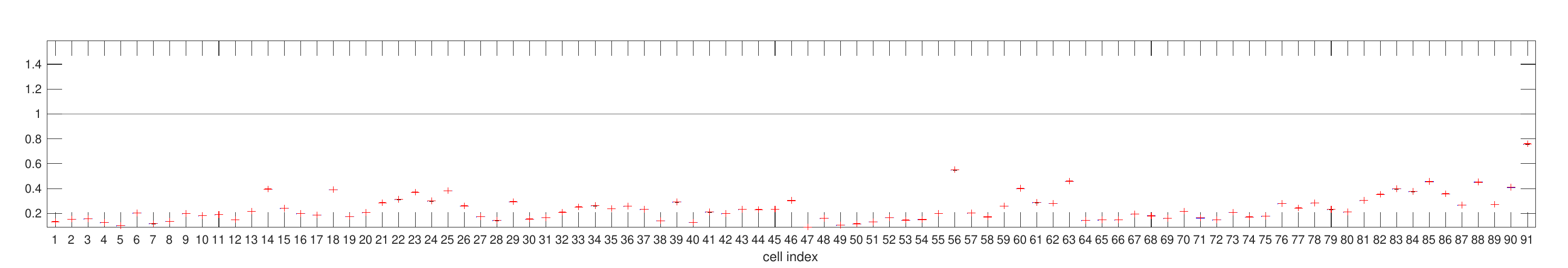}
\end{subfigure}%
\\
\vspace{0.1cm}
\begin{subfigure}{1\textwidth}
  \centering
   \caption{GLS, diagonal $V$}   \vspace{-0.2cm}
  \includegraphics[width = 5in]{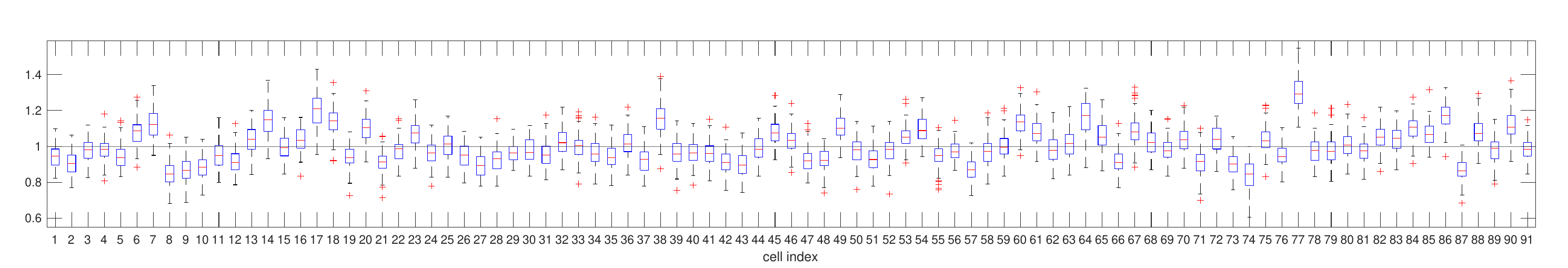}
\end{subfigure}%
\\
\vspace{0.1cm}
\begin{subfigure}{1\textwidth}
  \centering
   \caption{GLS, block-diagonal\ $V$}   \vspace{-0.2cm}
  \includegraphics[width = 5in]{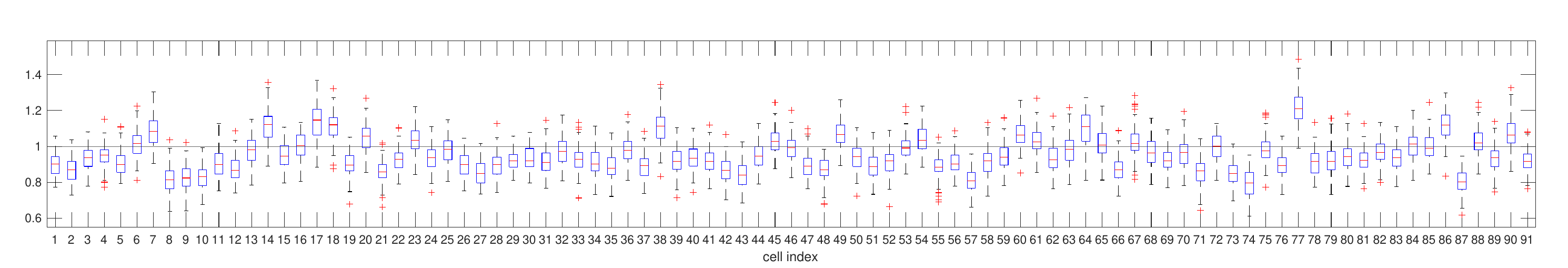}
\end{subfigure}%
\vspace{-0.1cm}
\caption{Boxplots of the ratios of the standard errors computed from the corresponding formula to the empirical standard deviation of $\hat\alpha_1$, with $100$ replications.
  Outliers are indicated by $+$'s, and a horizontal line marks $1$ (estimated standard deviation coincides with empirical standard deviation).
  Note that the vertical scale of panel (a) is different from (b) and (c), since OLS severely under-estimates standard errors.   Also, for OLS the variability for a given cell across simulations is dramatically smaller than variability between cells, which makes outliers the only easily visible part of the boxplots.   
}
\label{fig8}
\end{figure}

For valid inference, we need not just an accurate point estimate, but also an accurate standard error.   A comparison of the standard error of $\hat{\alpha}_1$ estimated using OLS and GLS to the empirical standard deviation of the estimated parameter $\F{sd}(\hat{\alpha}_1)$ is shown in  Figure~\ref{fig8}, with boxplots of the ratio $\mbox{estimated s.e.}/ \F{sd}(\hat{\alpha}_1)$ for the three estimators.  The  three rows in  Figure~\ref{fig8} show the results from OLS, GLS with a diagonal $V$, and GLS with a block-diagonal $V$, from (a) to (c). Ideally, these boxplots should be centered around 1, but the OLS ratios are much smaller than 1, though also the most stable;  the GLS standard errors, on the other hand, are much closer to the truth and also more variable themselves.   This shows that OLS severely under-estimates standard errors by assuming independence, while GLS leads to honest inference.
Careful scrutiny of Figure~\ref{fig8} reveals that the block-diagonal $V$ method \emph{slightly} underestimates standard errors, perhaps due to the slight overfitting resulting from its unneeded (in this setting) additional flexibility: the mean ratio of the diagonal $V$ method is $1.001$, whereas the mean ratio of the block-diagonal $V$ method is $0.951$.

\begin{figure}[H]
\centering
\begin{subfigure}{1\textwidth}
  \centering
      \caption{OLS}  \vspace{-0.2cm}
  \includegraphics[width = 5in]{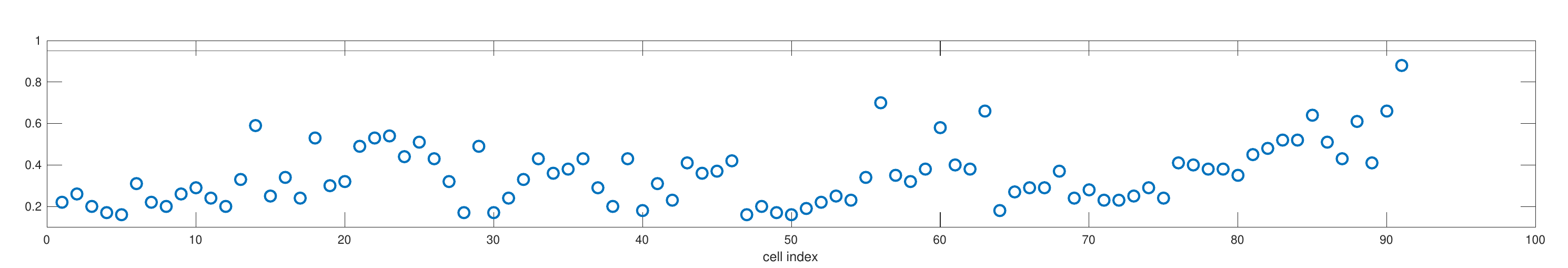}
\end{subfigure}%
\\
\vspace{0.1cm}
\begin{subfigure}{1\textwidth}
  \centering
   \caption{GLS, diagonal $V$}   \vspace{-0.2cm}
  \includegraphics[width = 5in]{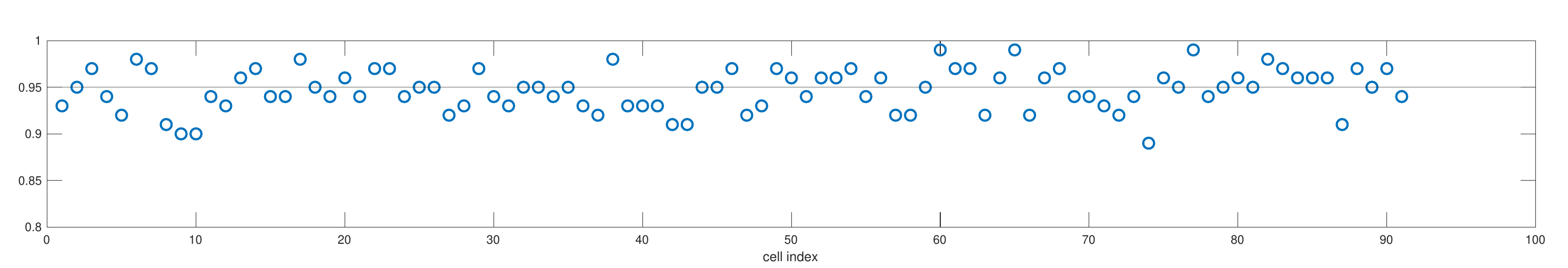}
\end{subfigure}%
\\
\vspace{0.1cm}
\begin{subfigure}{1\textwidth}
  \centering
   \caption{GLS, block-diagonal\ $V$}   \vspace{-0.2cm}
  \includegraphics[width = 5in]{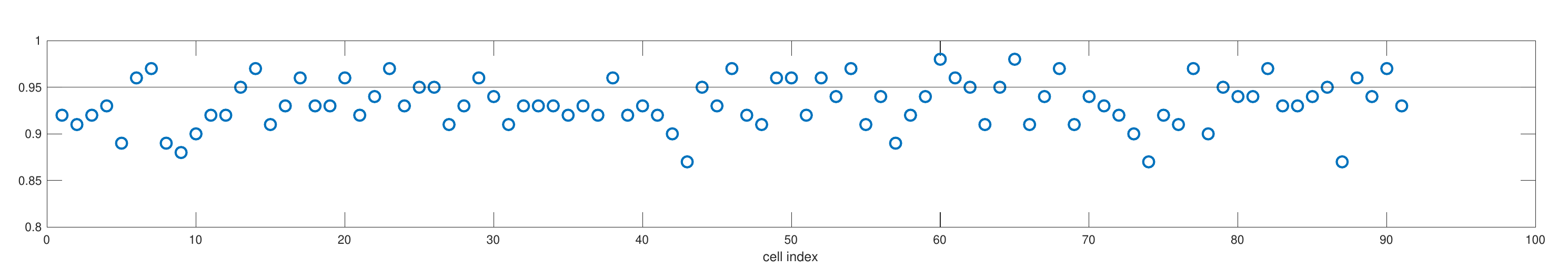}
\end{subfigure}%
\vspace{-0.1cm}
\caption{Coverage of the confidence intervals, $[\hat{\alpha}_1-1.96*s.e.(\hat{\alpha}_1), \hat{\alpha}_1+1.96*s.e.(\hat{\alpha}_1)]$, with $100$ simulations. The horizontal line corresponds to $95\%$ coverage.
  Note that the vertical scale in panel (a) is different from panels (b) and (c) due to poor coverage by OLS.}
\label{fig9}
\end{figure}

We also compared coverage rates of 95\%  confidence intervals for $\alpha_1$, defined by $[\hat{\alpha}_1-1.96*s.e.(\hat{\alpha}_1), \hat{\alpha}_1+1.96*s.e.(\hat{\alpha}_1)]$ for the three estimators, as shown in Figure~\ref{fig9}.  As we can expect from Figure~\ref{fig8}, OLS confidence intervals have poor coverage, but both GLS methods give coverage close to $95\%$: averaging across network cells, coverage is approximately $94.7\%$  for diagonal $V$ and  $93.2\%$ for block-diagonal $V$.   
Because coverage for the GLS method is reasonably close to nominal, for these two methods we further computed the false positive rate (FPR) and the true positive rate (TPR) to assess the size and power of the procedure, respectively.
The diagonal $V$ method had an FPR of approximately $0.054$ and a TPR of approximately $0.369$, while the block-diagonal $V$ method had an FPR of approximately $0.070$ and a TPR of approximately $0.336$.   

While we do not claim that this data-based simulation gives us a good estimate of just how far off the standard errors are in real data, 
we argue that it does show the errors will be unrealistically small if the model is fitted with OLS.   This is generally what we expect when the observations are positively correlated, and with many aspects of the simulated data matching the real data, this simulation gives us some idea of the differences between OLS and GLS performance that can be expected to arise in the application.  


\subsection{Validity under global null}
\label{sec:hc-only-analysis}
Next, we evaluate the distribution of $p$-values obtained by our method under the null hypothesis of no difference between two populations.   We use only the 70 healthy controls and split them randomly into two groups of 35.   In this case, 
$p$-values obtained by fitting the model should be uniformly distributed if the model is performing valid inference.  

\begin{figure}[H]
\centering
\begin{subfigure}{0.33\textwidth}
  \centering
      \caption{OLS}  \vspace{-0.2cm}
  \includegraphics[width = 1.6in]{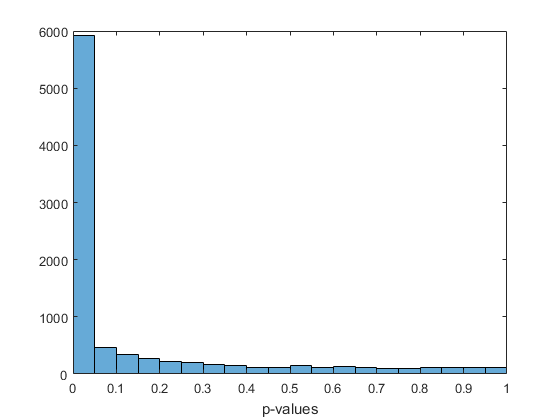}
\end{subfigure}%
\begin{subfigure}{0.33\textwidth}
  \centering
    \caption{GLS, diagonal $V$} \vspace{-0.2cm}
  \includegraphics[width = 1.6in]{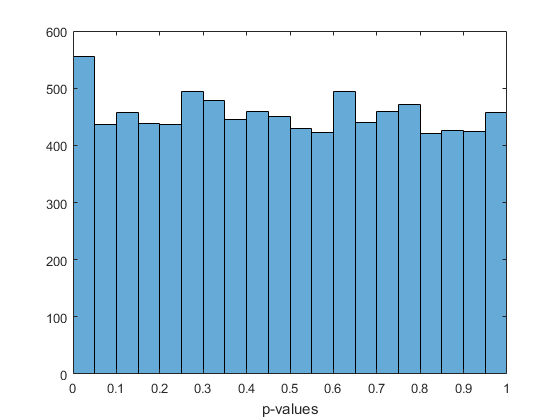}
\end{subfigure}%
\begin{subfigure}{0.33\textwidth}
  \centering
     \caption{GLS, block-diagonal\ $V$}   \vspace{-0.2cm}
  \includegraphics[width = 1.6in]{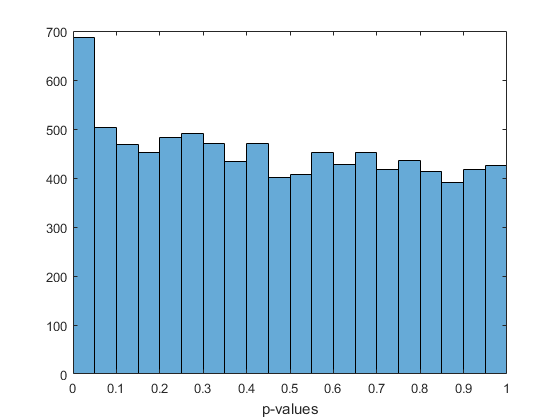}
\end{subfigure}
\vspace{-0.1cm}
\caption{Distribution of $p$-values under the null hypothesis.}
\label{fig10}
\end{figure}

We repeat the random splits into two groups 100 times, fitting our model and computing $p$-values for the hypotheses $\alpha_1^{(a, b)} = 0$ every time, resulting in a total 9100 $p$-values (91 cells $\times$ 100 repetitions). Figure~\ref{fig10} shows the histogram of these 9100 $p$-values from OLS and GLS.     The GLS histograms look close to the uniform, whereas the OLS histogram has a large number of small $p$-values, and approximately 2/3 are less than 0.05.  After applying the Benjamini-Hochberg multiple testing correction \citep{Benjamini1995} to control FDR at $5\%$ (as we do in our application), the OLS method rejects the null hypothesis, on average, for $56.7$ cells (out of $91$) in each replication of the simulation.  In contrast, the models with diagonal and block-diagonal $V$ reject, respectively,  $0.23$ and $0.37$ out of 91 hypotheses on average. Even with the more conservative Bonferroni correction, OLS still rejects $41$ cells on average, whereas the rates for the two GLS methods are $0.14$ and $0.28$.  To quantify the comparison of the distribution of the $p$-values to the uniform, we apply a Kolmogorov-Smirnov test comparing the $p$-values for a given cell across the 100 replications to the uniform distribution.     This gives us 91 resulting $p$-values for each method shown in Figure \ref{fig10}, which we again compare to the uniform distribution by a Kolmogorov-Smirnov test (note that because there may be some dependence across cells, the $p$-value obtained from this procedure is not strictly valid, but still serves as a concise data summary).  As may be anticipated from Figure \ref{fig10}, applying this procedure gives a $p$-value of 0 (to machine precision) for OLS, 0.6098 for GLS with a diagonal $V$, and 0.01653 for GLS with a block-diagonal $V$.

\subsection{Parameter estimation in full dataset}
\label{sec:cobre-fitting}

\begin{figure}[H]
\centering
\begin{subfigure}{0.5\textwidth}
  \centering
      \caption{Healthy, diagonal $V$}  \vspace{-1.5cm}
  \includegraphics[width = 2.4in]{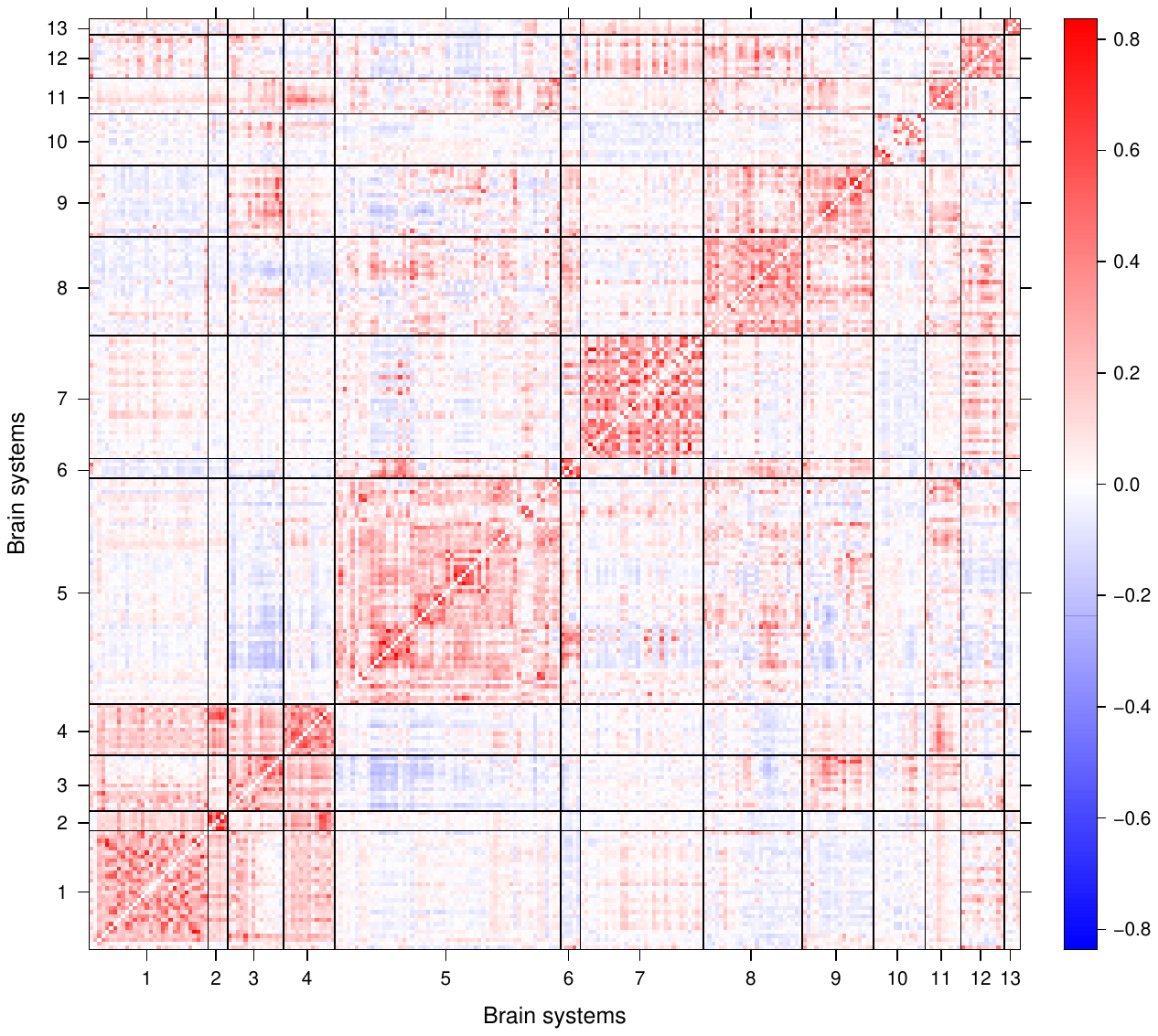}
     \label{fig:M1_diag}
\end{subfigure}%
\begin{subfigure}{0.5\textwidth}
  \centering
    \caption{Schizophrenia, diagonal $V$} \vspace{-1.5cm}
  \includegraphics[width = 2.4in]{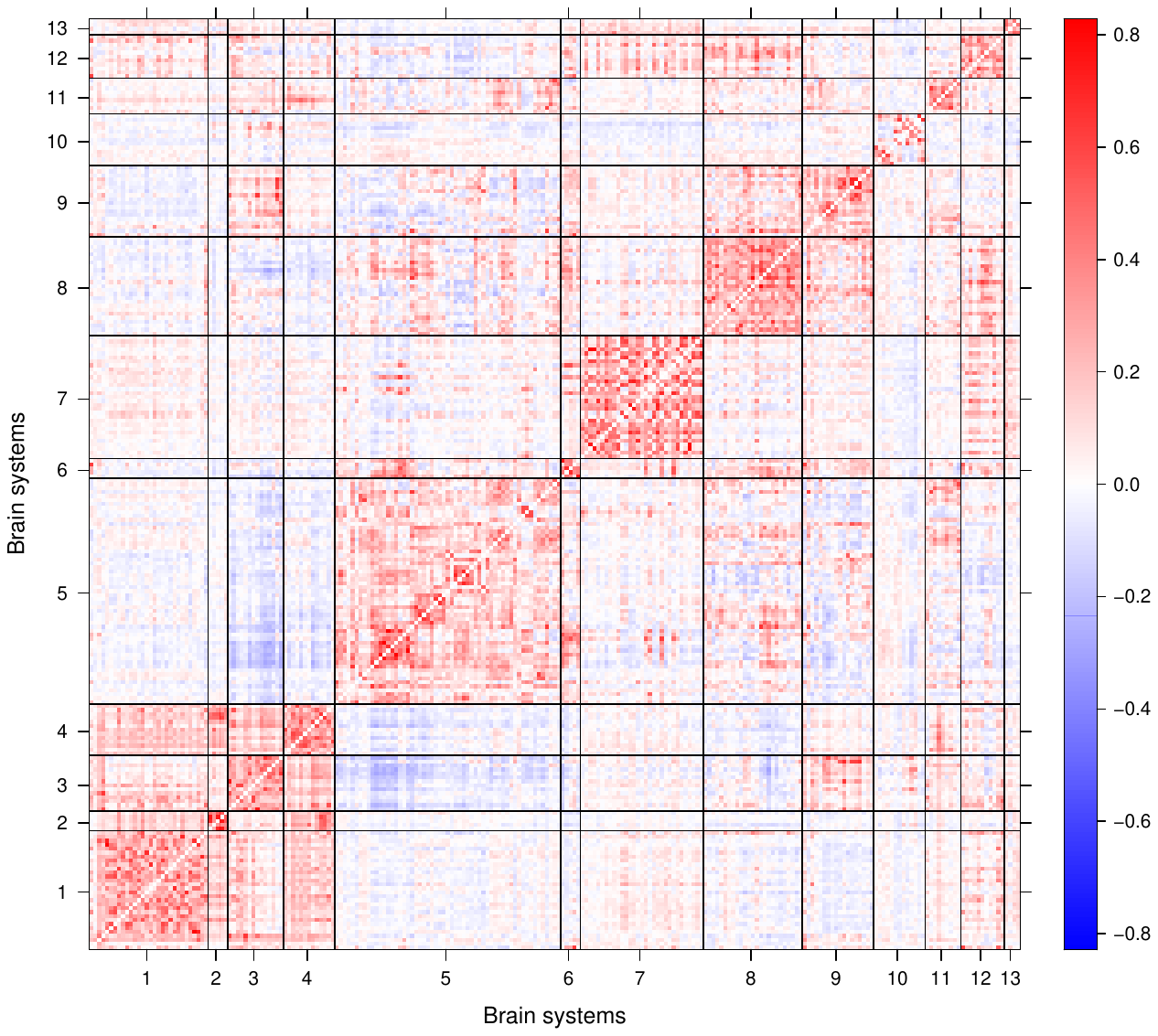}
  \label{fig:M2_diag}
\end{subfigure}%
\\
\vspace{-1cm}
\begin{subfigure}{0.5\textwidth}
  \centering
   \caption{Healthy, block-diagonal\ $V$}   \vspace{-1.5cm}
  \includegraphics[width = 2.4in]{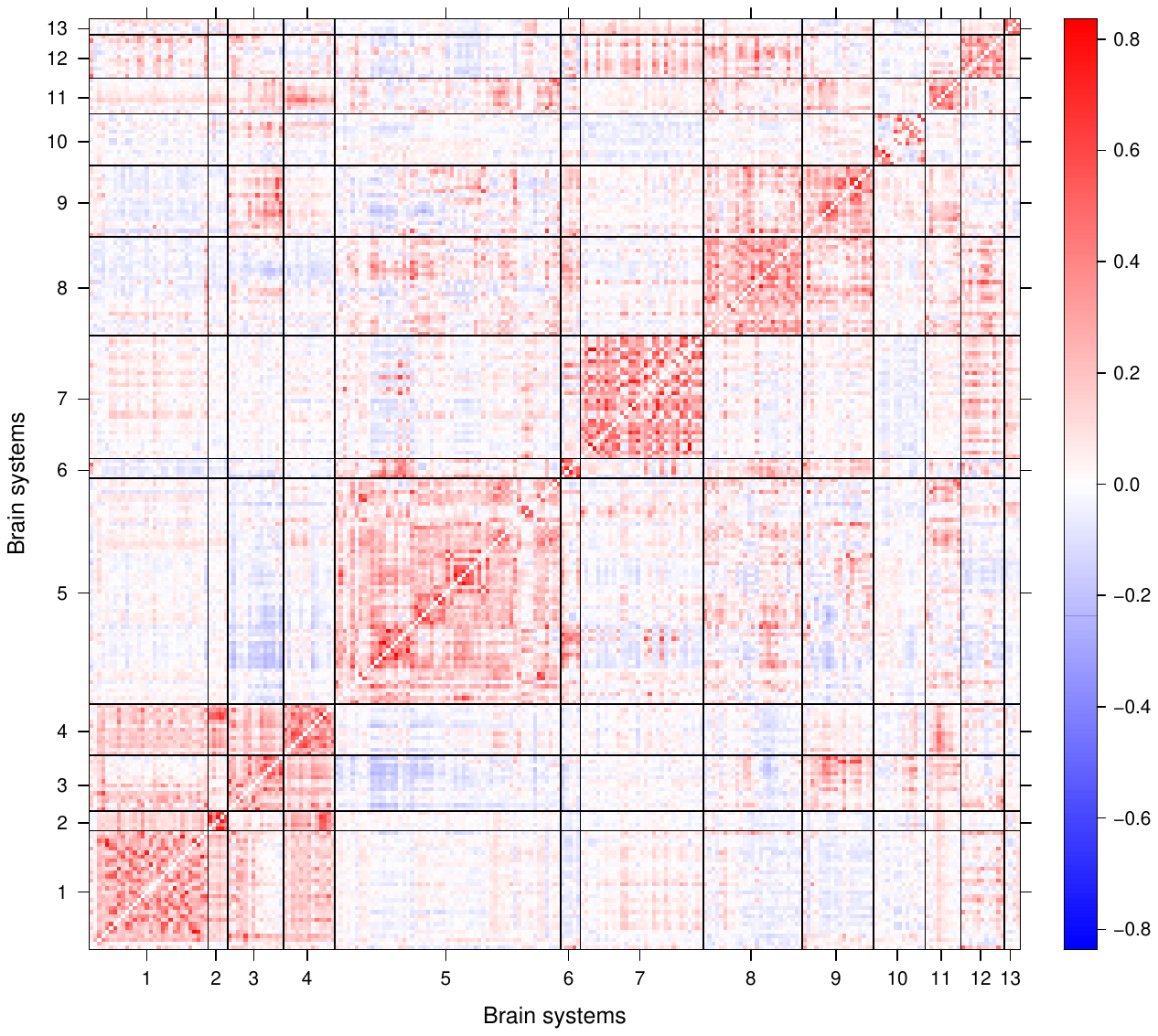}
  \label{fig:M1_blkd}
\end{subfigure}%
\begin{subfigure}{0.5\textwidth}
  \centering
     \caption{Schizophrenia, block-diagonal\ $V$}   \vspace{-1.5cm}
  \includegraphics[width = 2.4in]{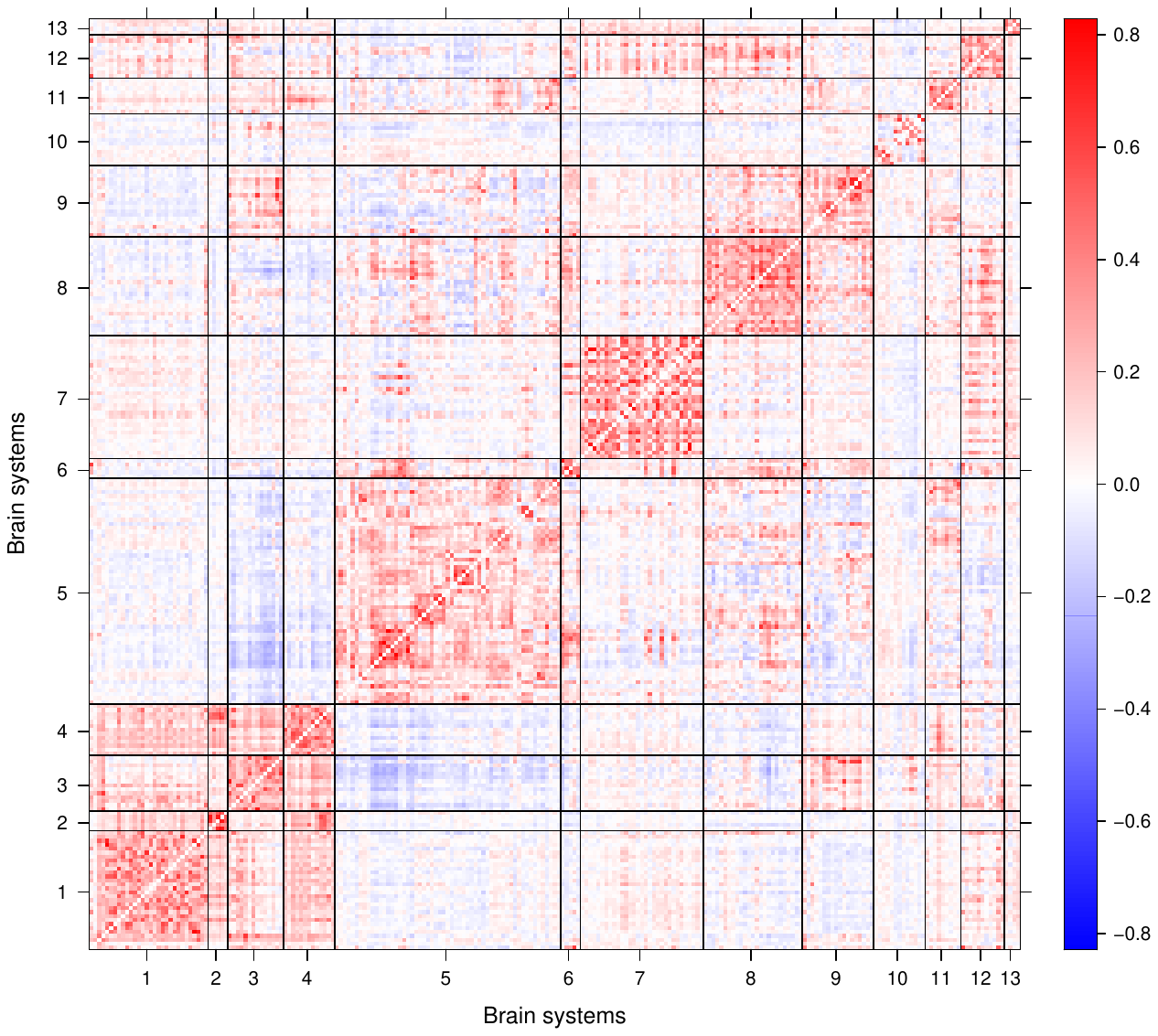}
  \label{fig:M2_blkd}
\end{subfigure}
\vspace{-1.2cm}
\caption{Estimates of the mean networks for healthy and schizophrenia patients. } 
\label{fig:means}
\end{figure}

We now fit model \eqref{eq:model_conditional} to the entire COBRE dataset, with disease status as a binary covariate.    Figure~\ref{fig:means} shows the estimated mean networks for the two populations, i.e., $\{\hat{\alpha}_0+\hat{\eta}_{i, 0}\}$ and $\{\hat{\alpha}_0+\hat{\alpha}_1+\hat{\eta}_{i, 0}+\hat{\eta}_{i, 1}\}$, with either diagonal or block-diagonal $V$.   The dominant structure in the means is the community structure, with stronger connectivity within each functional system, which is expected.

\begin{figure}[H]
\centering
\begin{subfigure}{0.5\textwidth}
  \centering
      \caption{Cell-level, diagonal $V$}  \vspace{-1.5cm}
  \includegraphics[width = 2.4in]{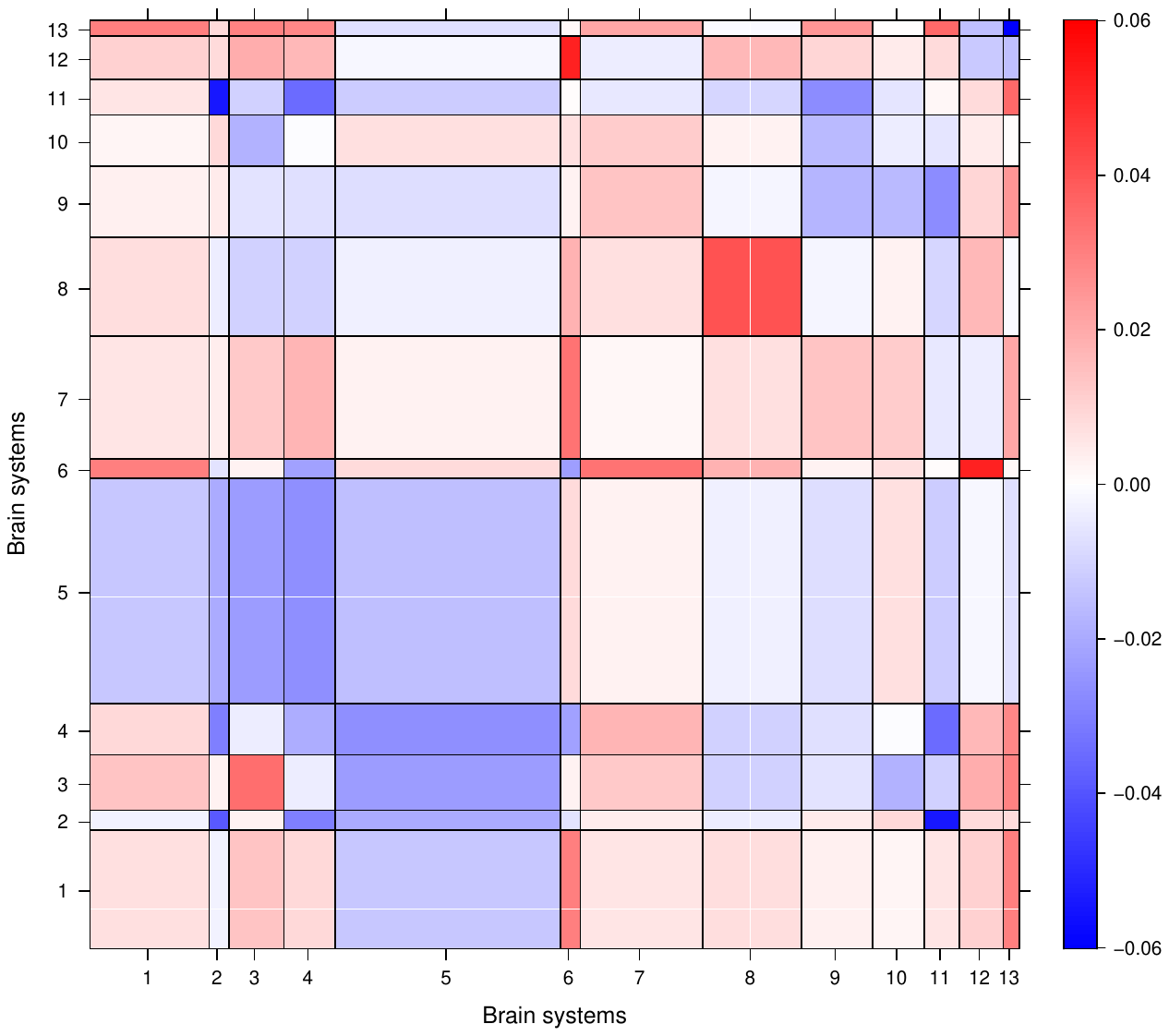}
     \label{fig:alpha_diag}
\end{subfigure}%
\begin{subfigure}{0.5\textwidth}
  \centering
    \caption{Edge-level, diagonal $V$} \vspace{-1.5cm}
  \includegraphics[width = 2.4in]{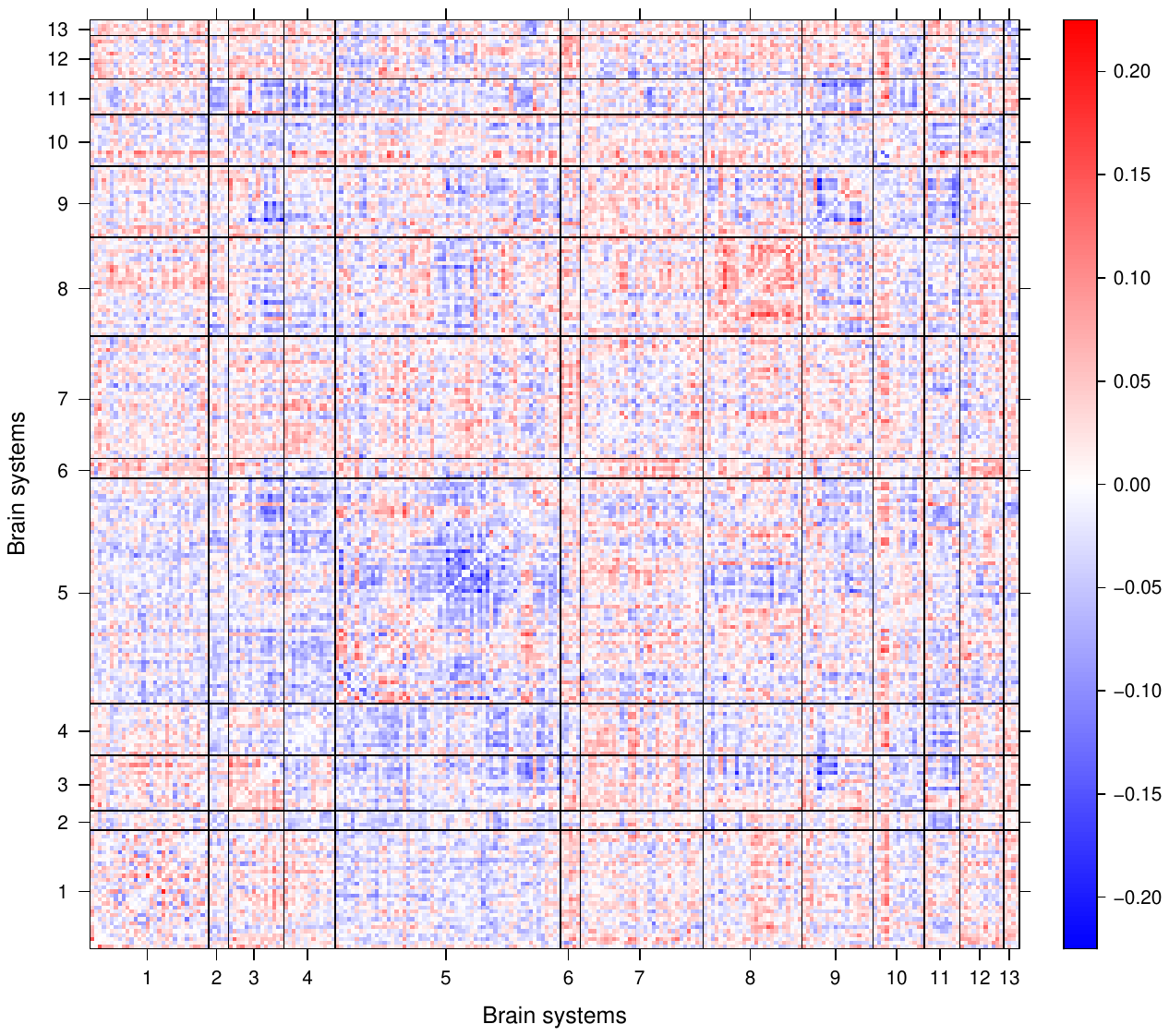}
  \label{fig:edge_diag}
\end{subfigure}%
\\
\vspace{-1cm}
\begin{subfigure}{0.5\textwidth}
  \centering
   \caption{Cell-level, block-diagonal\ $V$}   \vspace{-1.5cm}
  \includegraphics[width = 2.4in]{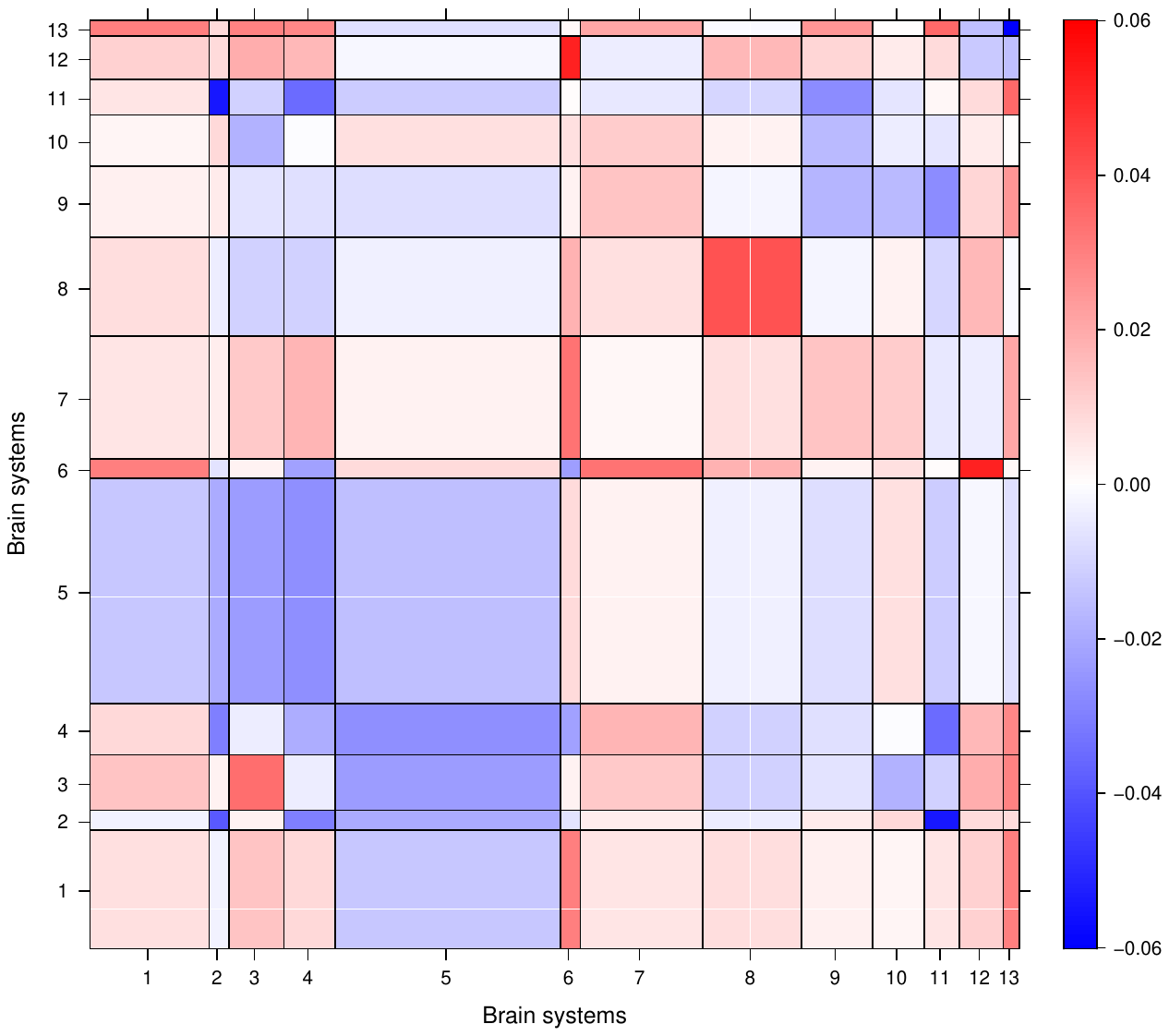}
  \label{fig:alpha_blkd}
\end{subfigure}%
\begin{subfigure}{0.5\textwidth}
  \centering
     \caption{Edge-level, block-diagonal\ $V$}   \vspace{-1.5cm}
  \includegraphics[width = 2.4in]{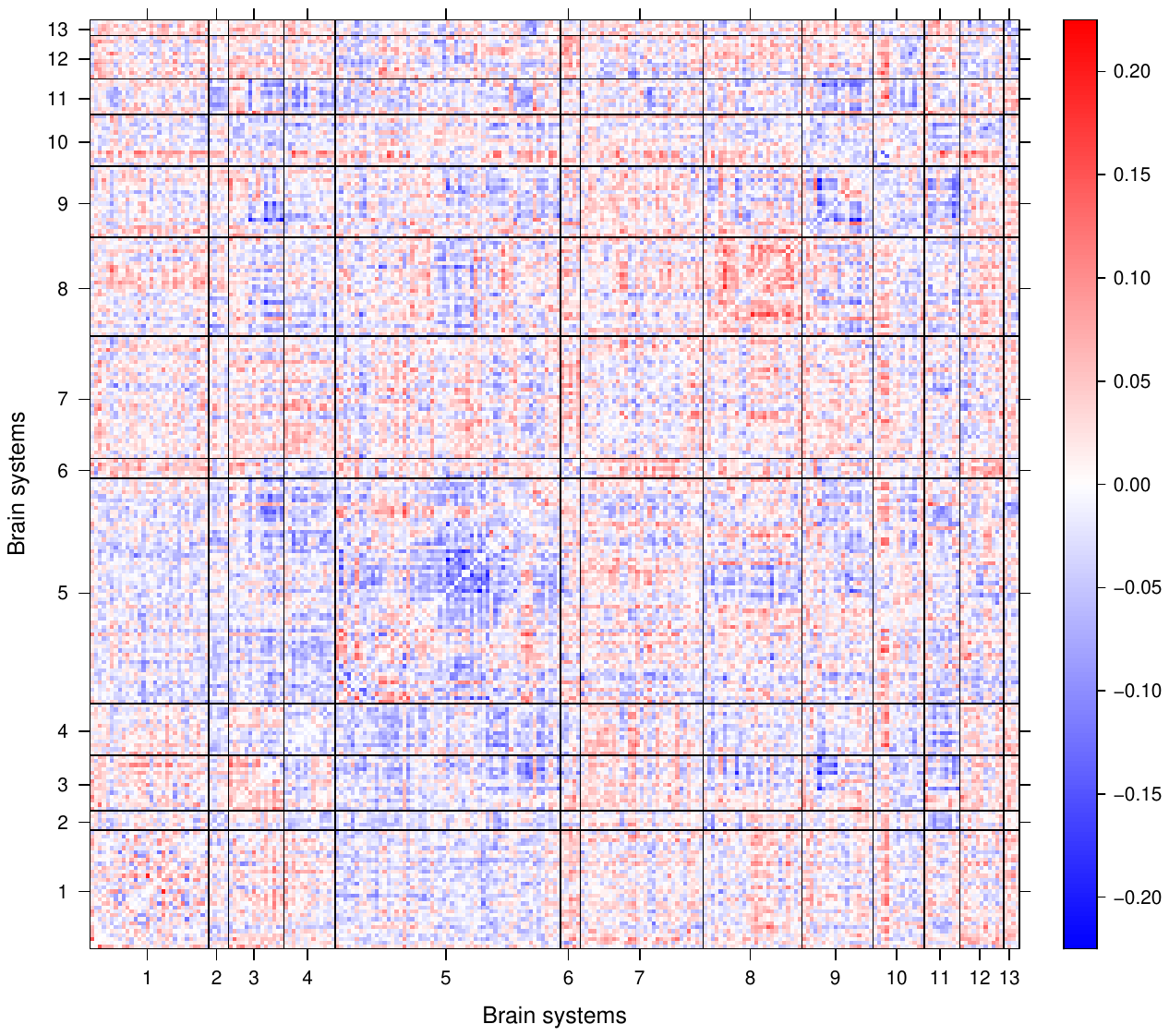}
  \label{fig:edge_blkd}
\end{subfigure}
\vspace{-1.2cm}
\caption{Estimates of differences between the two populations, healthy and schizophrenia, with diagonal and block-diagonal $V$.   Blue represents larger mean for control group and red represents larger mean for schizophrenia group.}
\label{fig:diff}
\end{figure}

Figure~\ref{fig:alpha_diag} and Figure~\ref{fig:alpha_blkd} show the cell-level differences in connectivity between the two populations, i.e.,  $\hat{\alpha}_1$.  Negative $\hat{\alpha}_1$ (blue) corresponds to higher values for controls,  and  positive $\hat{\alpha}_1$ (red) for schizophrenic patients.   The higher control values in cell $(9,9)$, corresponding to lower connectivity within the salience system for schizophrenic patients, match a previously reported dysfunction in schizophrenia \citep{Palaniyappan2013}.   Schizophrenia effects on the frontal and parietal brain regions (system 8) also have been reported \citep{VandenHeuvel2010}.

Figure~\ref{fig:edge_diag} and Figure~\ref{fig:edge_blkd} show differences between the two populations at the edge level, i.e., $(\{\hat{\alpha}_0+\hat{\alpha}_1+\hat{\eta}_{i, 0}+\hat{\eta}_{i, 1}\})-(\{\hat{\alpha}_0+\hat{\eta}_{i, 0}\})$ for each edge $i$. The edge effects are quite heterogeneous, especially for the large cells, such as (5,5). The heterogeneous cell effects suggest that we do need to include edge-level effects, and in fact interpreting cell effects without the edge effects may lead to misleading results.

\subsection{Hypothesis testing for group comparisons}
\label{sec:cobre-inference}

\begin{figure}[H]
\centering
\begin{subfigure}{0.5\textwidth}
  \centering
  \caption{BH correction, diagonal $V$}   \vspace{-1.5cm}
  \includegraphics[width = 2.4in]{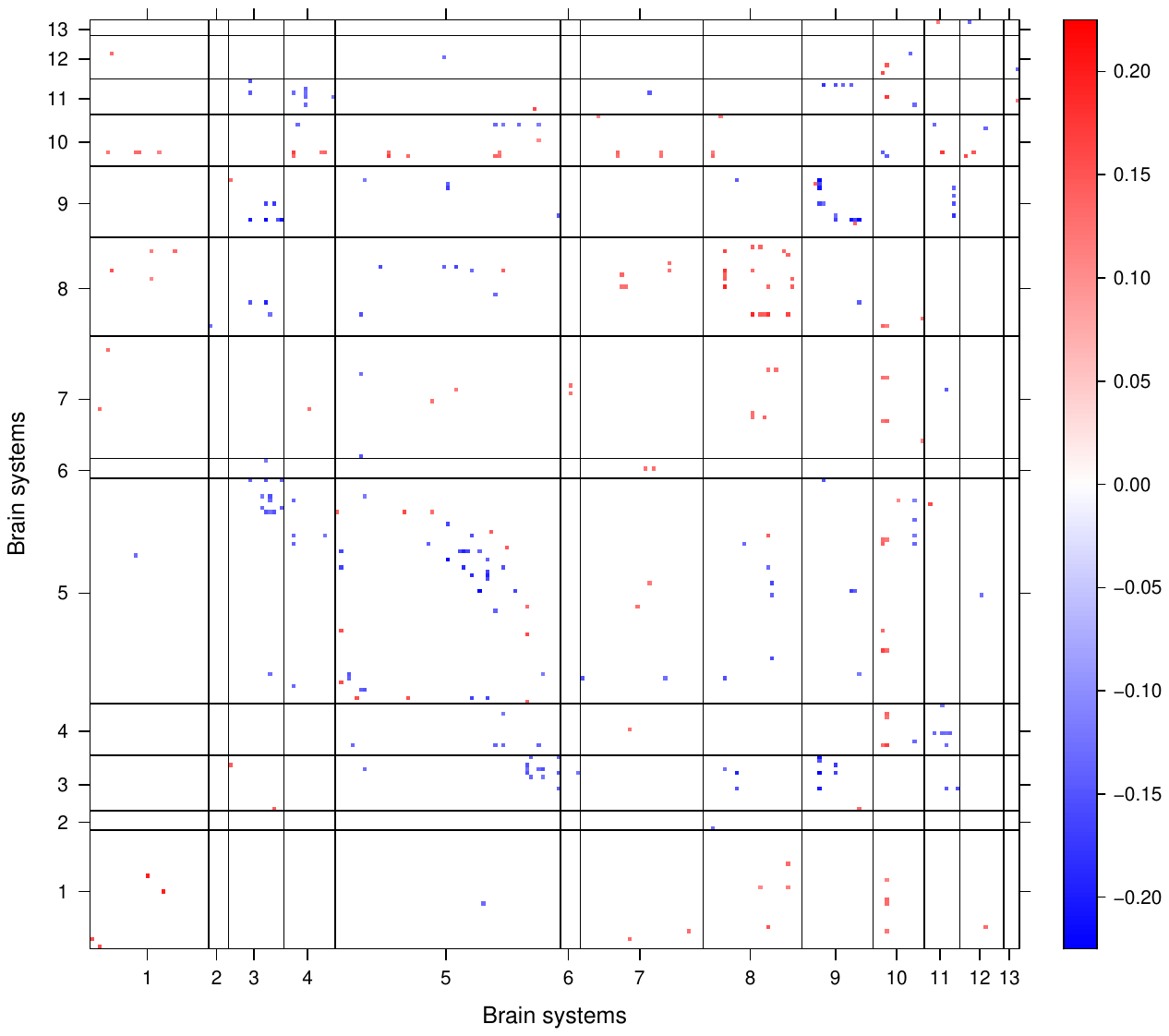}
\end{subfigure}%
\begin{subfigure}{0.5\textwidth}
  \centering
  \caption{BH correction, block-diagonal\ $V$}   \vspace{-1.5cm}
  \includegraphics[width = 2.4in]{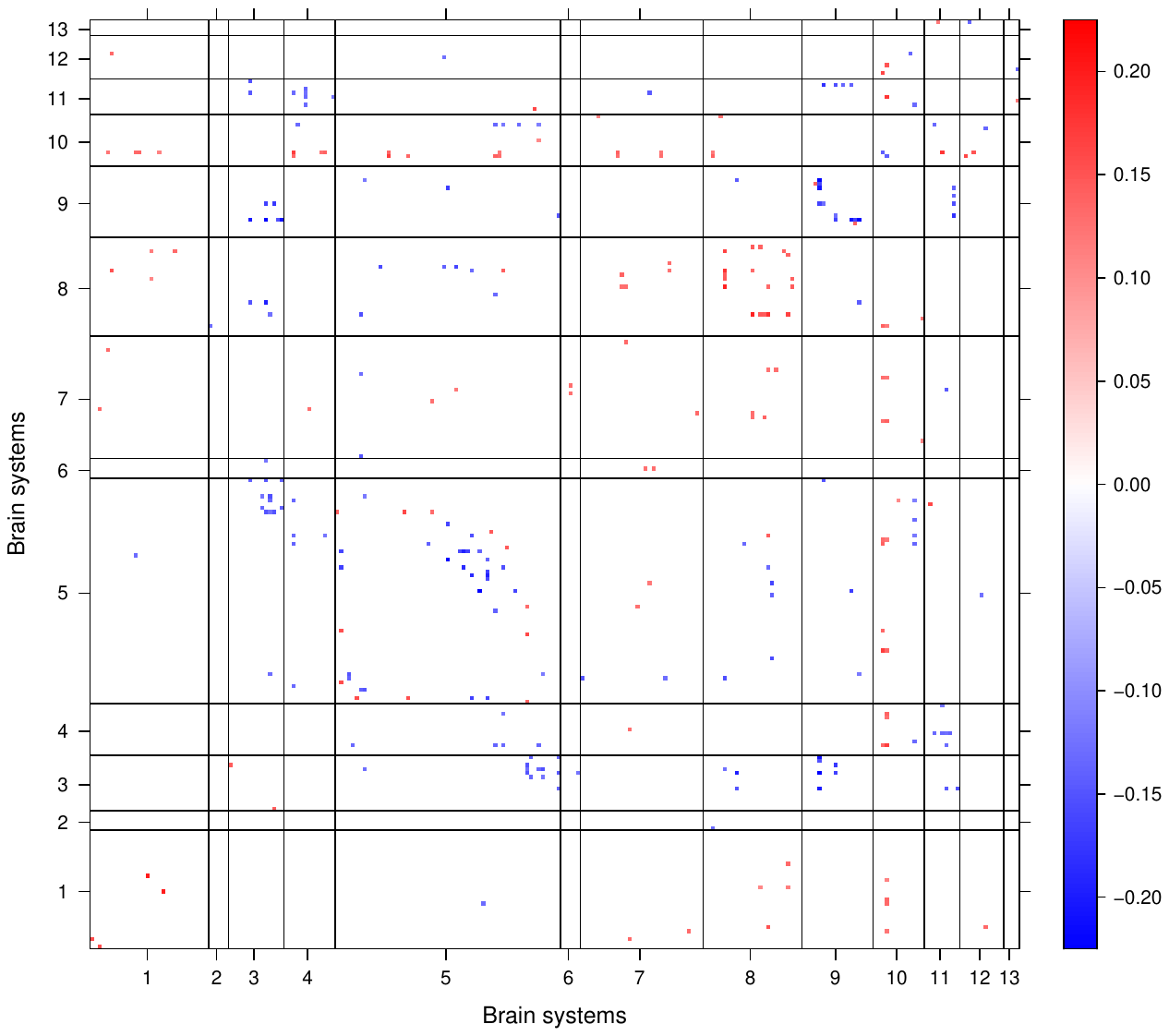}
\end{subfigure}
\vspace{-1.2cm}
\caption{Estimates of edge-level difference between healthy vs. schizophrenia for significant edges at $5\%$ significance level (after Benjamini-Hochberg correction).}
\label{fig:sigedge}
\end{figure}

To assess the differences between groups more formally, we perform hypothesis tests.    First, consider testing the edge-level effects, with the null $H_{0,i}:  \alpha_1+\eta_{i, 1} =0$ versus alternative $H_{1,i}: \alpha_1+\eta_{i, 1} \neq 0$ for every edge $i$.
Results with the Benjamini-Hochberg multiple testing correction, controlling FDR at $5\%$, are presented in Figure~\ref{fig:sigedge}.   Both diagonal and block-diagonal $V$ give very similar results with 150 and 149 significant edge-level differences, respectively;  148 edges out of these two sets are the same.

We chose the Benjamini-Hochberg correction 
primarily because it does well on power, but other choices are possible, including methods that control family-wise error rate (FWER) such as Bonferroni's, Holm's \citep{Holm1979}, and Hochberg's \citep{Hochberg1988}, or the Benjamini-Yekutieli method for controlling FDR under dependency \citep{Benjamini2001}.  Applying the Benjamini-Yekutieli procedure to our data yielded only 22 significant edges for both diagonal $V$ and block-diagonal $V$; while this procedure is valid under dependency, it is known to be conservative.

\begin{figure}[H]
\centering
\begin{subfigure}{0.5\textwidth}
  \centering
      \caption{Left hemisphere, diagonal $V$}  \vspace{-0.25cm}
  \includegraphics[width = 2.6in]{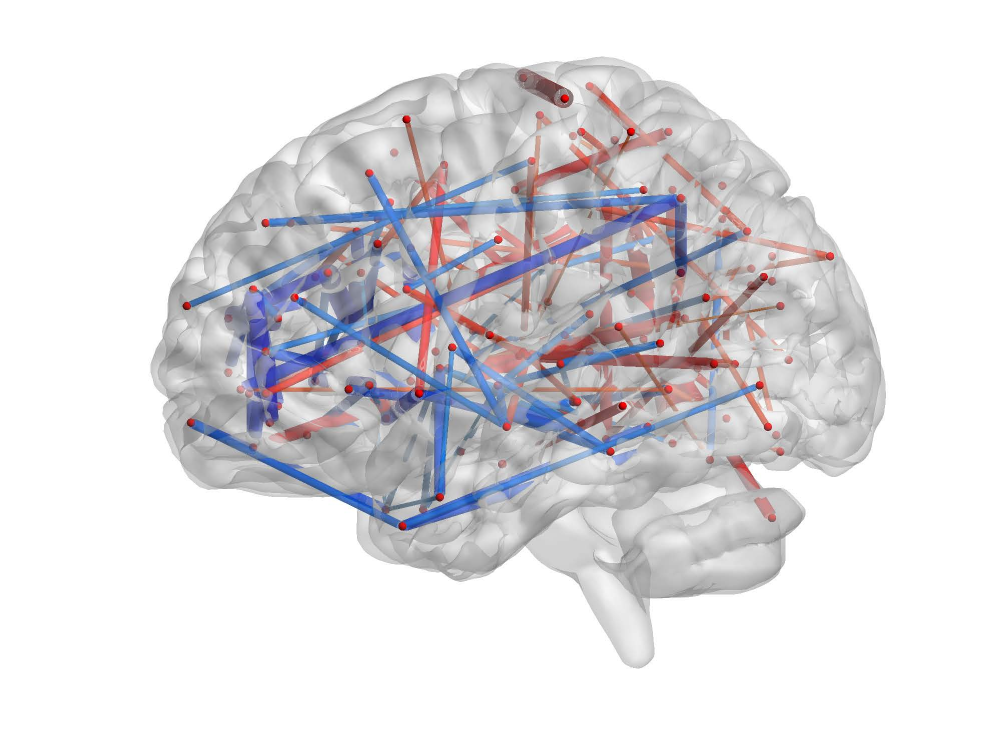}
\end{subfigure}%
\begin{subfigure}{0.5\textwidth}
  \centering
    \caption{Right hemisphere, diagonal $V$} \vspace{-0.25cm}
  \includegraphics[width = 2.6in]{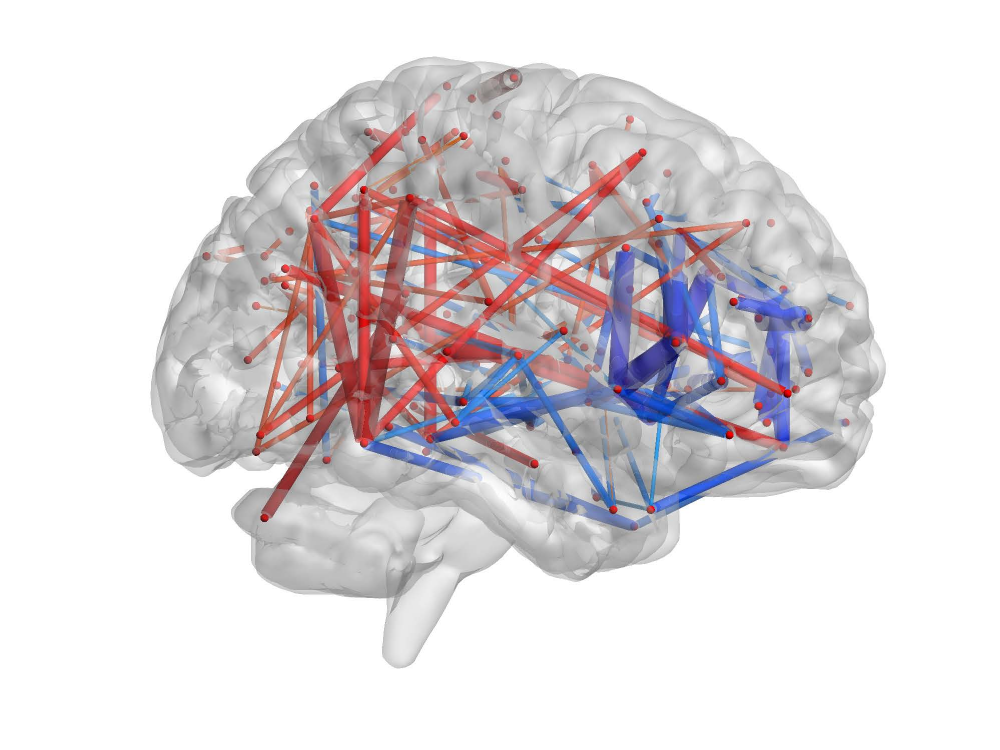}
\end{subfigure}%
\\
\vspace{-0.5cm}
\begin{subfigure}{0.5\textwidth}
  \centering
   \caption{Left hemisphere, block-diagonal\ $V$}   \vspace{-0.25cm}
  \includegraphics[width = 2.6in]{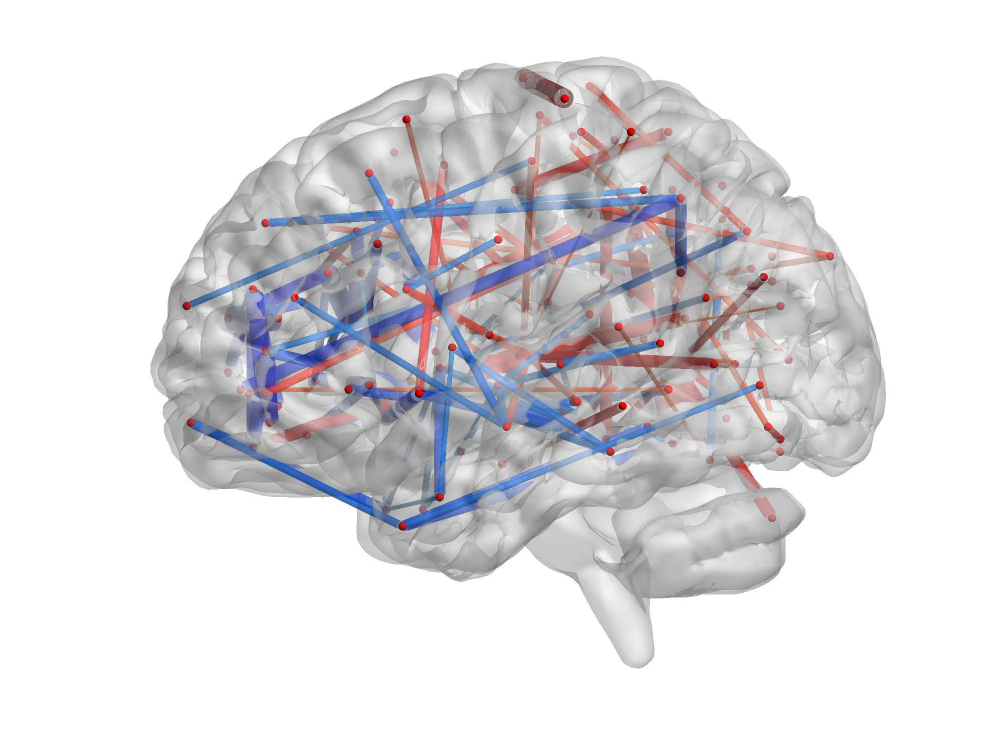}
\end{subfigure}%
\begin{subfigure}{0.5\textwidth}
  \centering
     \caption{Right hemisphere, block-diagonal\ $V$}   \vspace{-0.25cm}
  \includegraphics[width = 2.6in]{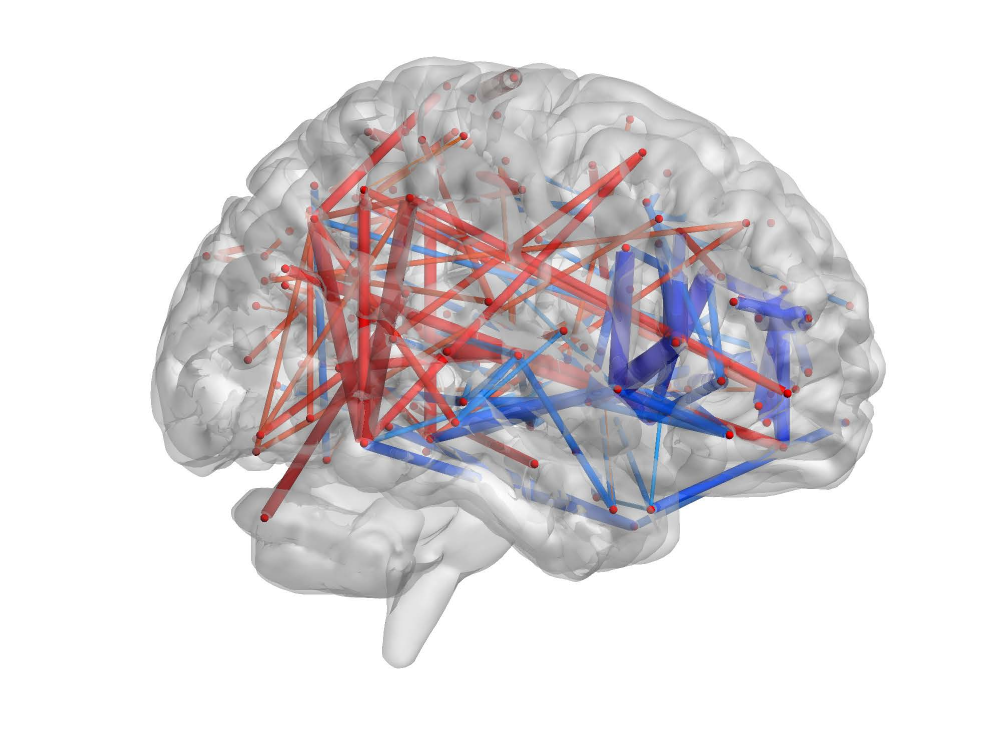}
\end{subfigure}
\vspace{-0.5cm}
\caption{Estimates of edge-level difference on the brain. Blue edges represent larger mean for healthy control and red edges represent larger mean for schizophrenia. Line width represents the magnitude of the difference. Figures generated using BrainNet Viewer \citep{Xia2013}.}
\label{fig:braindiff}
\end{figure}

Figure~\ref{fig:braindiff} shows the positive and negative edge differences plotted on the brain.    Our results generally agree with previous findings on schizophrenia, including a disconnection between the frontal and the temporal cortices \citep{Friston1995, Bullmore2011} and occipito-temporal disconnections \citep{Zalesky2010}.    Figure~\ref{fig:braindiff} clearly shows the asymmetric connectivity difference between healthy control and schizophrenia for left and right hemispheres, which is also aligned with previous studies \citep{Mitchell2005, Angrilli2009, Ribolsi2014}. 

\begin{figure}[H]
\centering
\begin{subfigure}{0.5\textwidth}
  \centering
      \caption{Diagonal $V$}  \vspace{-1.5cm}
  \includegraphics[width = 2.4in]{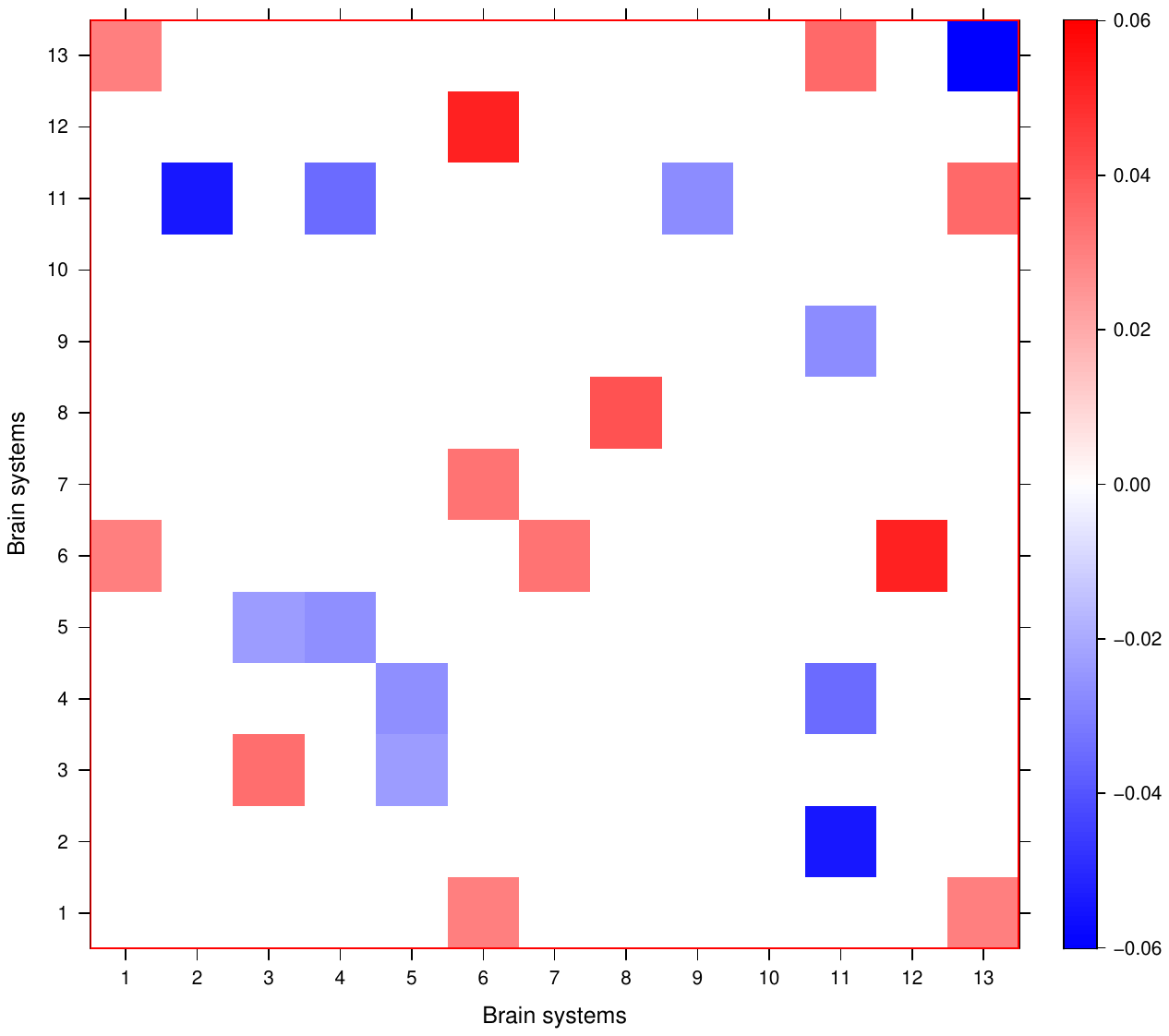}
\end{subfigure}%
\begin{subfigure}{0.5\textwidth}
  \centering
    \caption{Block-diagonal\ $V$} \vspace{-1.5cm}
  \includegraphics[width = 2.4in]{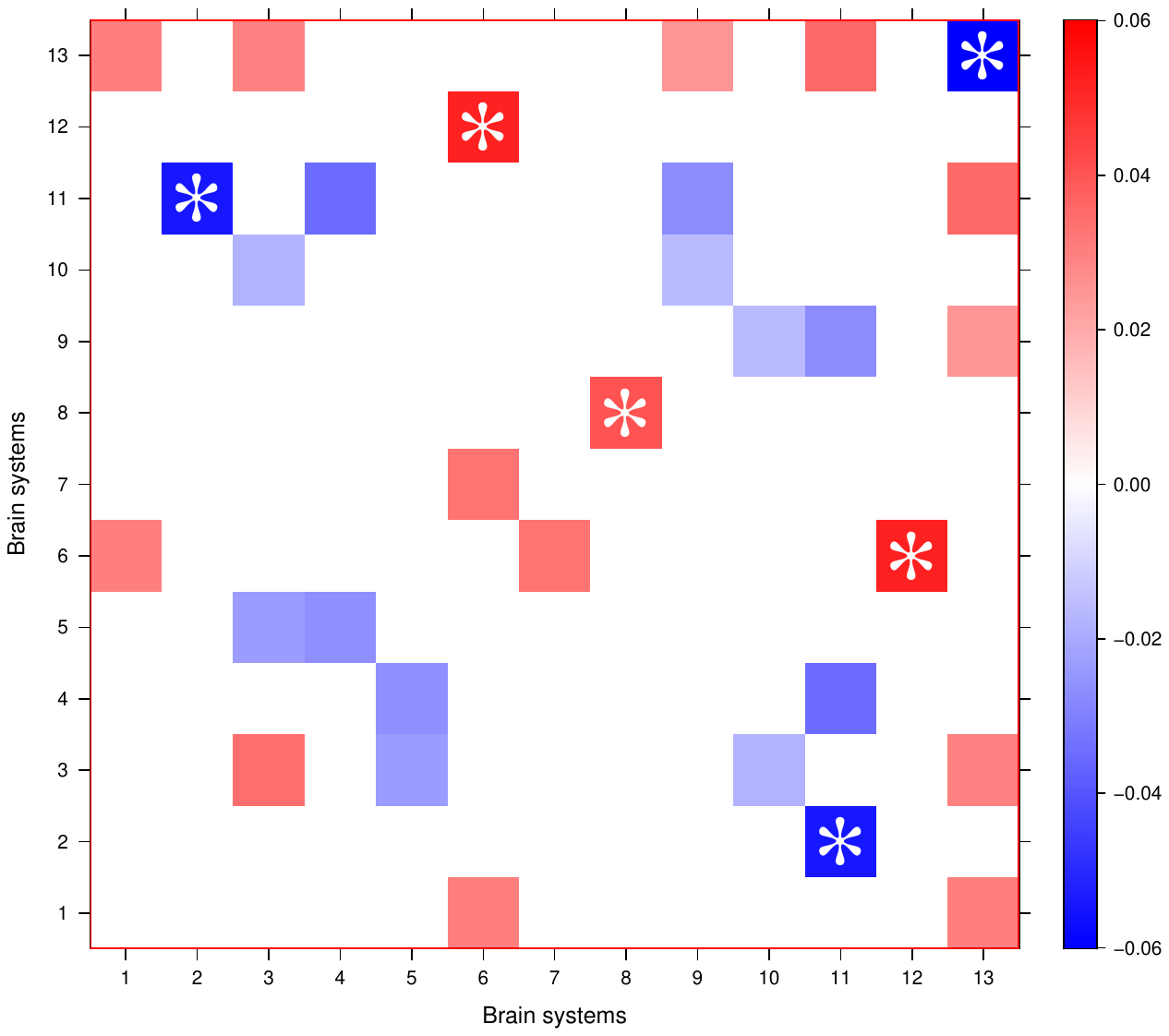}
\end{subfigure}
\vspace{-1.2cm}
\caption{Values of $\hat{\alpha}_1^{(a, b)}$ for cells $(a,b)$ with $\hat\alpha_1^{(a, b)}$ significant at $5\%$ before multiple testing correction.   Cells retaining significance after the Benjamini-Hochberg correction are marked with an asterisk (*).  With diagonal $V$, there are no significant cells after correction.} 
\label{fig:sigcell}
\end{figure}

We next test the cell-level hypotheses  $H_{0}^{(a, b)}: \alpha_1^{(a, b)}=0$ versus $H_{1}^{(a, b)}: \alpha_1^{(a, b)}\neq 0$ for every network cell  $(a,b)$. Results, both prior to and after correction for multiple testing, are presented in Figure~\ref{fig:sigcell}.   At the cell level, the increase in connectivity within system 8, the fronto-parietal task control system, was associated with the lowest $p$-value, both for diagonal and block-diagonal $V$, as shown in Figure~\ref{fig:sigcell}.   This is consistent with a previous study \citep{Venkataraman2012}, which reported increased connectivity between parietal and frontal regions.

\begin{figure}[H]
\centering
\includegraphics[width = 5in]{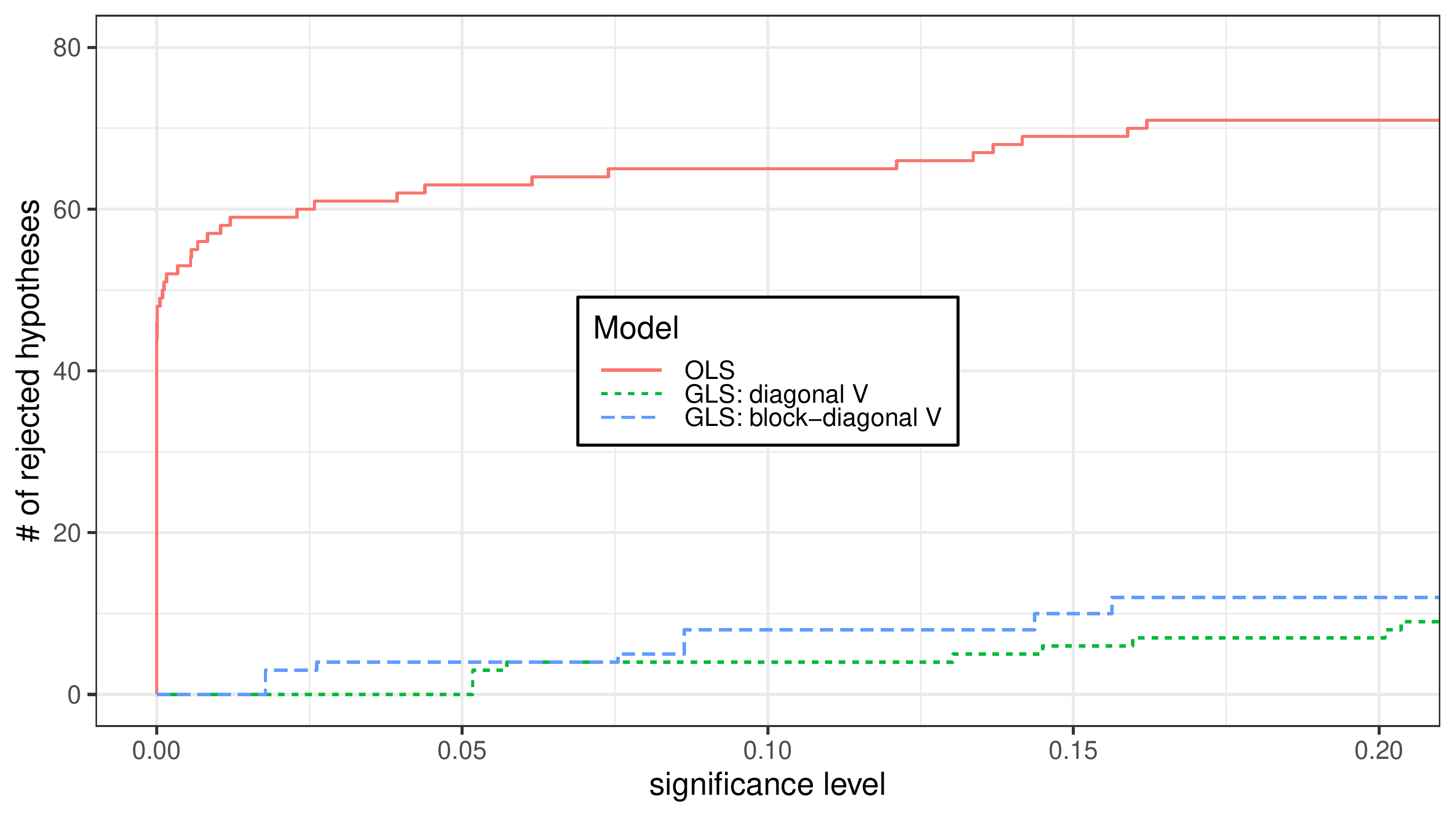}
\caption{Number of rejected hypotheses as a function of significance level with a Benjamini-Hochberg multiple testing correction. Red: OLS; green: GLS with diagonal $V$; blue: GLS with block-diagonal $V$.}
\label{fig13}
\end{figure}

With varying significance level, we also look at the number of rejected hypotheses (out of 91) using OLS and GLS (with diagonal/ block-diagonal $V$) after Benjamini-Hochberg correction.
From Figure~\ref{fig13}, we can see that OLS rejects many more hypotheses (over half even at the smallest significance value) than either version of GLS.  Our simulation results suggest this is due to the under-estimation of standard errors with OLS.   The GLS results are more realistic:  from Figure~\ref{fig13}, we can see that at $5\%$ significance level, GLS with diagonal $V$ does not find anything, and GLS with a block-diagonal $V$ finds four significant cells, $(8,8)$, $(2,11)$, $(6,12)$, and $(13,13)$; see Figure~\ref{fig:sigcell}. 
 
\begin{figure}[H]
\centering
\includegraphics[width = 0.48\textwidth]{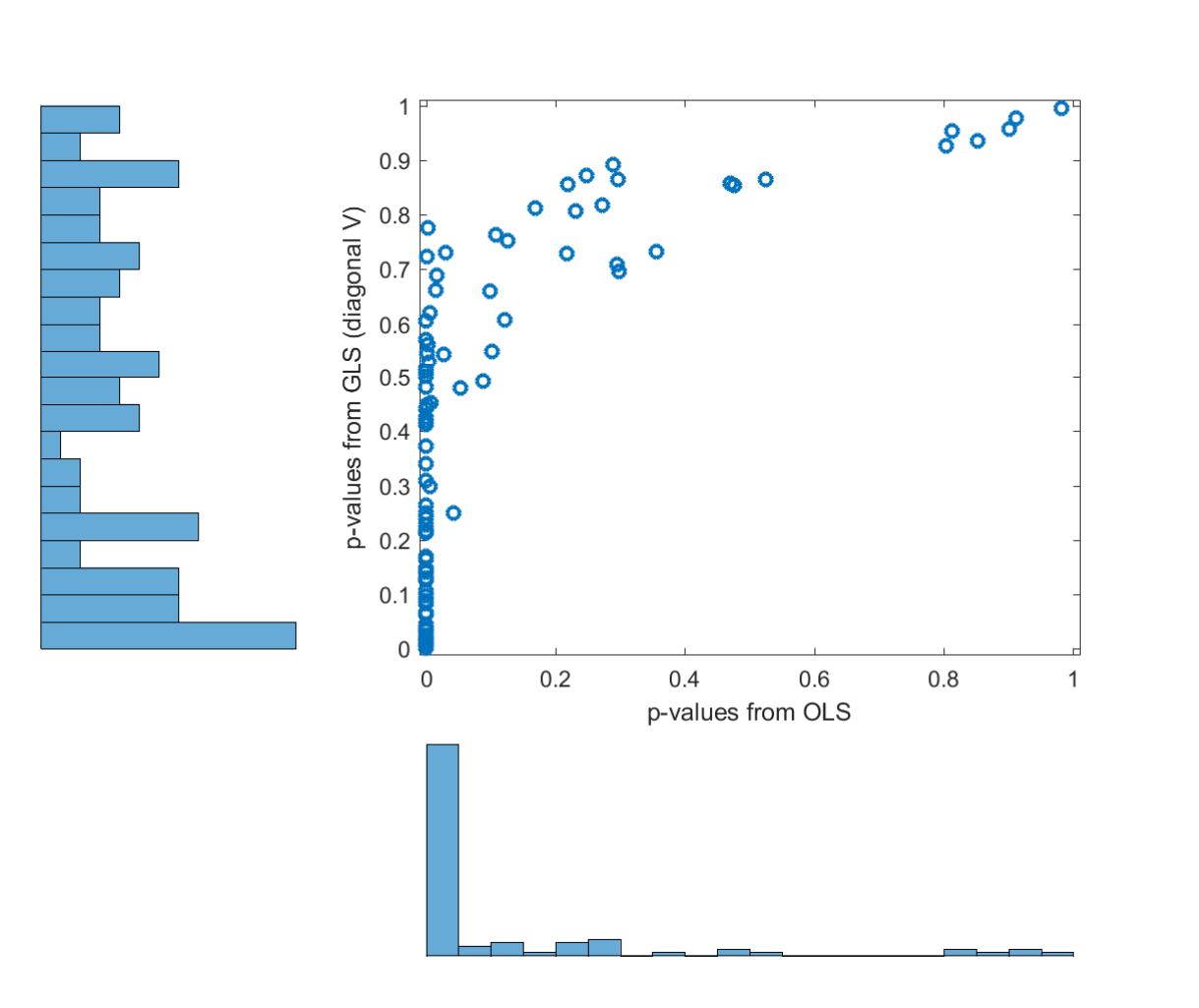}
\includegraphics[width = 0.48\textwidth]{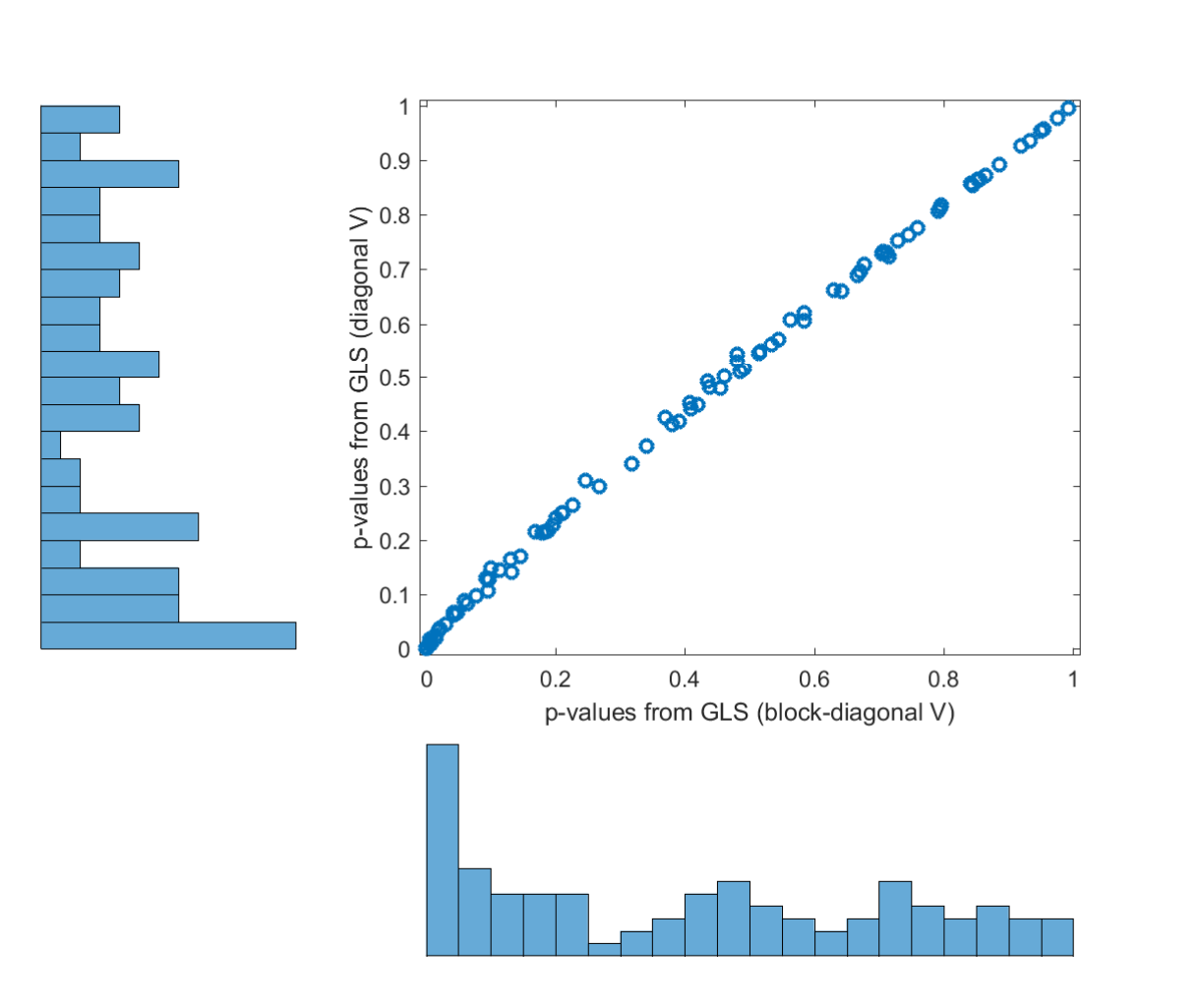}
\caption{Scatter plot of raw $p$-values. Left: OLS versus GLS with block-diagonal $V$; right: GLS with diagonal $V$ versus GLS with block-diagonal $V$. Histograms depict the marginal distribution of $p$-values for each setting.}
\label{fig14}
\end{figure}

The distributions of raw $p$-values from OLS and GLS are shown in Figure~\ref{fig14}. Consistent with the simulation findings, OLS produces a large number of small $p$-values, and the ordering of $p$-values is also fairly different between OLS and GLS.   On the other hand, the $p$-values from GLS with diagonal and block-diagonal $V$ agree closely, and the cells with the lowest $p$-values match. Top five cells ordered by $p$-values are shown in Table~\ref{table2}. 

\begin{table}[H]
\footnotesize
\centering
\begin{tabular}{ccccc}
  \hline
$p$-value rank  & network cell & diagonal $V$ & block-diagonal $V$  & adjusted significance level \\ 
  \hline
1 & (8,8) & 8.9e-04 & 3.0e-04 & 5.5e-04 \\ 
2 & (6,12) & 0.0012 &  5.1e-04 & 0.0011 \\ 
3 & (2,11) & 0.0017 & 5.9e-04 & 0.0016 \\
4 & (13,13) & 0.0025 & 0.0012 & 0.0022 \\
5 & (4,5) & 0.0072 &  0.0041 & 0.0027 \\ 
   \hline
\end{tabular}
\caption{Five most significant cells from GLS methods. The ordering of the cells obtained by two GLS versions is the same. Last three columns show the $p$-values from GLS methods and the Benjamini-Hochberg adjusted significance level, i.e., $\frac{0.05 k}{91}$ for $k = 1 \ldots 5$.} 
\label{table2}
\end{table}

\subsection{Comparison with \citet{xiaMultiscaleNetworkRegression2020}}
\label{sec:cobre-compare-xia}

\begin{figure}
  \centering
  \includegraphics[width = 0.48\textwidth]{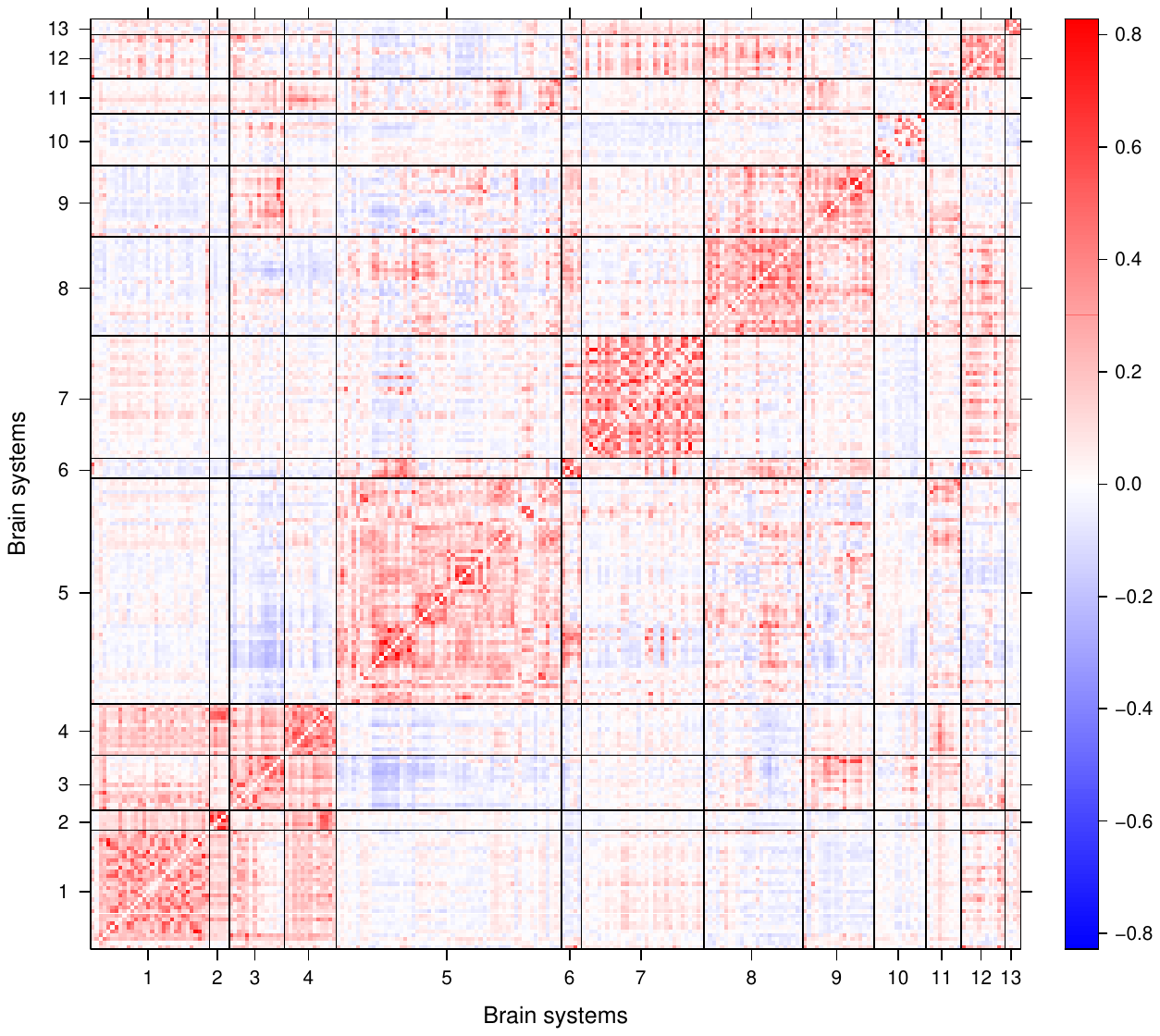}
  \includegraphics[width = 0.48\textwidth]{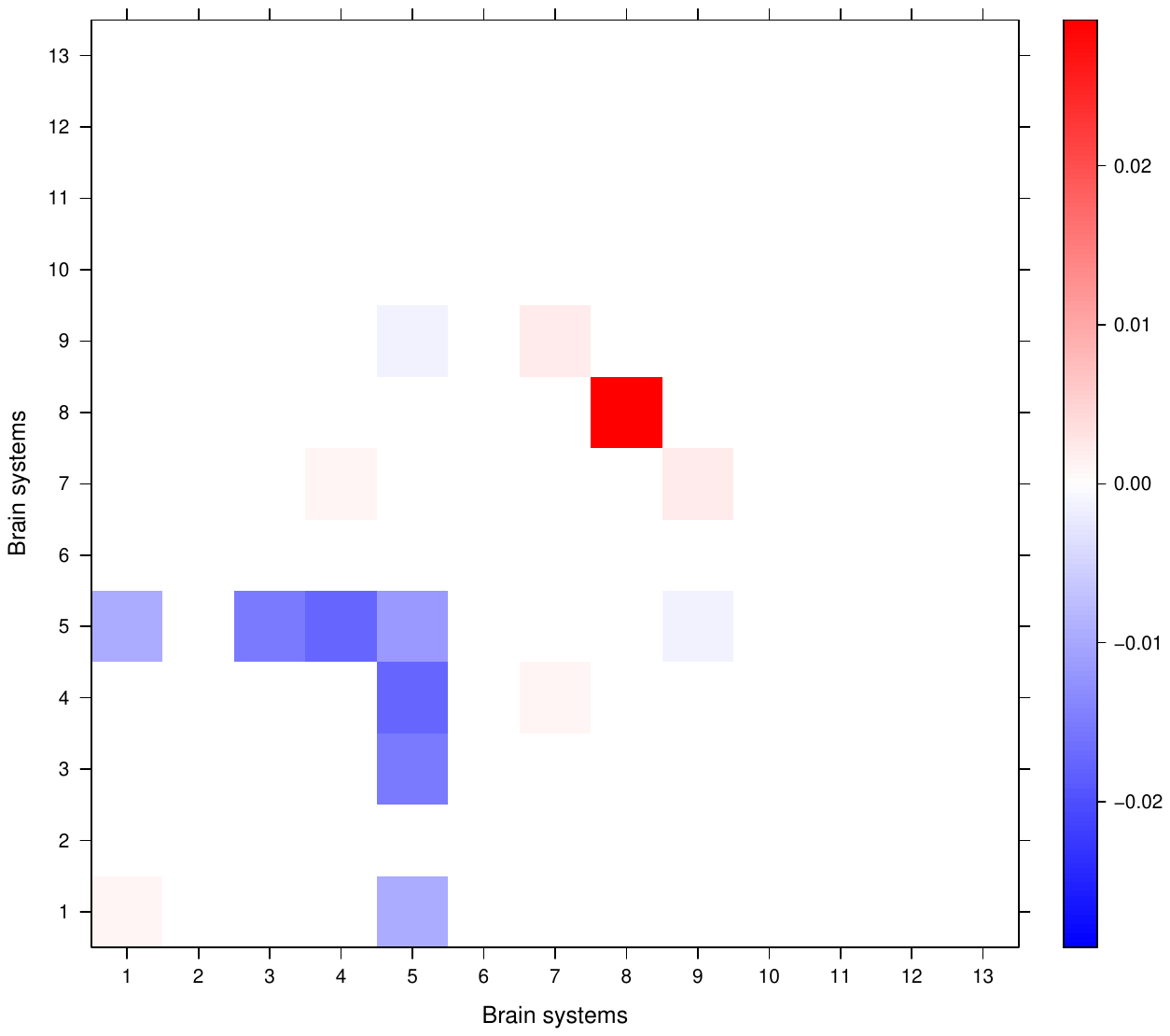}
  \caption{Results of applying the method of \citet{xiaMultiscaleNetworkRegression2020}.
  Left: the intercept $\Theta$;  right: the cell effects $\alpha$.}
  \label{fig:xia-results}
\end{figure}

A related approach that can be applied in our setting was proposed by \citet{xiaMultiscaleNetworkRegression2020}, who fit a model of the relationship between phenotypes and brain connectivity using penalized least squares.     Adjusting their notation to match ours, they solve the optimization problem 
\begin{equation}
\label{eq:xia-objective}
  \operatorname*{minimize}_{\Theta, \alpha}
  \sum_{m=1}^N \lVert A^{(m)} - \Theta - d_m Z \alpha Z^{T} \rVert_F^2  +
  \lambda_1 \lVert \Theta \rVert_{\star} +
  \lambda_2 \lVert \alpha \rVert_1,
\end{equation}
where $\Theta \in \mathbb{R}^{n \times n}$ is the intercept, $\alpha \in \mathbb{R}^{K \times K}$ is a matrix of  cell effects, $d_m$ is a binary indicator for schizophrenia diagnosis, $\lVert \Theta \rVert_{\star}$ is the nuclear norm of $\Theta$ (i.e., the sum of the singular values), $\lVert \alpha \rVert_1$ is the sum of the absolute values of the entries of the matrix $\alpha$, and $\lambda_1$ and $\lambda_2$ are tuning parameters.
We applied this method to the COBRE data using code available from \citet{xiaMultiscaleNetworkRegression2020}.
We selected tuning parameters by searching over a grid, with each parameter taking possible values in the set $\left\{ e^{-10}, e^{-9}, \ldots, e^{10} \right\}$, and computing the average loss from five-fold cross-validation at each point.
We then refitted the model with the best-performing parameters ($\lambda_1 = e^2, \lambda_2 = e^6$); we visualize both $\Theta$ and $\alpha$ in Figure \ref{fig:xia-results}.

The intercept $\Theta$ appears qualitatively similar to that estimated by our method (see left column of Figure \ref{fig:means}).
Since this method induces sparsity in the cell effects $\alpha$, we compare the cells it selected as non-zero to the cells our method identified as statistically significant (see Figure \ref{fig:sigcell}).
While there are some overlaps between the two sets of results (e.g., cells involving Brain System 5, the Default mode system), overall the patterns are not especially similar.
These dissimilarities can largely be explained, however, by considering the objective function \eqref{eq:xia-objective}, which does not adjust for the relative size of the network cells induced by the varying functional system  sizes.
For example, the cell $(5,5)$ comprises 1653 edges whereas the cell $(13,13)$ has only 6 edges.
Since the penalty on $\alpha$ does not adjust for this, the model tends to select the largest cells.
Whether or not this is a desirable property depends upon the question being asked;  in the present setting we are more interested in knowing which cells show group differences as opposed to just minimizing the fitted error of the model.  
Another important distinction is that the method of \citet{xiaMultiscaleNetworkRegression2020} does not provide inference, so it is more difficult to assign significance to the pattern of discovered effects.
Our approach, in contrast, provides $p$-values which can be then appropriately corrected for multiple comparisons.



%% file: sections/discussion.tex
\section{Discussion}
\label{sec:discussion}

We introduced a new framework for modeling multiple brain networks with the goal of taking network structure into account when assessing covariate effects, in modeling both the mean and the variance.   The choice of a linear mixed effects model allowed for a simple interpretable decomposition into cell-level and edge-level effects while accounting for  individual variation.    In an important departure from typical network models, we allowed for reasonably general dependency between a very large number of edges by allowing a general variance structure in the linear model.   While our application focused on a single binary disease status covariate, the method can be applied to a general linear model in subject-level covariates.  

Our empirical results suggest that without a variance structure that allows for some edge dependence, standard errors are severely underestimated, producing unreliable and misleadingly small $p$-values.    Modeling the variance with a network community structure, on the other hand, results in accurate standard errors, as shown both on synthetic data and on the analysis of healthy subjects from the COBRE data split into two parts at random.   While the analysis comparing schizophrenia patients and healthy controls has no ground truth, our results confirm important earlier findings, such as a disconnection between frontal and temporal cortices in schizophrenia  \citep{Friston1995, Bullmore2011}. 

Previously proposed methods for comparing brain networks tend to perform standard two-sample inference on either global graph summaries or vectorized edge weights.
While the global graph summaries do account for network structure, they tend to collapse a lot of information, and generally do not do as well in classification tasks \citep{Jesus2017}.
Vectorized edge weights, on the other hand, preserve all the information, but do not take advantage of the network structure.
In contrast, our method allows inference both at the level of cell means (as depicted e.g., in Figure~\ref{fig:sigcell}) and at the level of individual edges (as depicted e.g., in Figure~\ref{fig:sigedge}), and it does so while accounting for edge dependence.
Importantly, it provides an interpretable model and in particular hypothesis testing for specific systems in the brain, which none of the previous methods can easily do.
System-level inference also greatly reduces the number of hypotheses to be tested compared to the massive-univariate approach (vectorized edge weights), which suggests our tests have more power after correcting for multiple testing.  
While some previous work \citep{xiaMultiscaleNetworkRegression2020} does directly consider system-level effects, it does not 
readily provide inference for these effects, and so results must generally be interpreted qualitatively.

One limitation of any approach that works with edge weights 
is that these are summary statistics drawn from a sequence of time courses acquired via fMRI; shorter time courses or higher levels of noise will limit the effectiveness of the method.   While we include individual edge random effects that can reflect this noise in edge weights, there is a limit to how useful the results can be in the presence of very noisy edge weights.   Another potential limitation of our method is using a pre-determined parcellation.   While one can learn a parcellation first, or even use the fitted edge effects to improve a given parcellation (e.g. split cells that show a lot of heterogeneity), validity of inference in the presence of such a model selection step is unclear.   Still, we believe developing graph-aware approaches that strike a balance between massively univariate but information-preserving methods and global summaries is a fruitful direction for future work on multiple network analysis, both for the brain imaging application and other applications involving multiplex networks, such as global trade networks (in multiple commodities), transportation networks (by different means of transport), social networks (with different types of connections), and many others.

